\documentclass[pdflatex,sn-nature]{sn-jnl}

\usepackage{lmodern}
\usepackage{graphicx}%
\usepackage{multirow}%
\usepackage{tabularx}
\usepackage{amsmath,amssymb,amsfonts}%
\usepackage{amsthm}%
\usepackage{mathrsfs}%
\usepackage[title]{appendix}%
\usepackage{xcolor}%
\usepackage{textcomp}%
\usepackage{manyfoot}%
\usepackage{booktabs}%
\usepackage{algorithm}%
\usepackage{algorithmicx}%
\usepackage{algpseudocode}%
\usepackage{listings}%
\usepackage{longtable}%
\usepackage{mathtools}
\usepackage{logicpuzzle}%
\usepackage{subfig}

\theoremstyle{thmstyleone}%

\theoremstyle{thmstyletwo}%

\theoremstyle{thmstylethree}%

\usepackage{xspace}
\usepackage[normalem]{ulem}
\newcommand{\eg}{e.g.,\xspace}
\newcommand{\idest}{i.e.,\xspace}

\usepackage{acronym}
\acrodef{QC}[QC]{Quantum Computing}
\acrodef{QA}[QA]{Quantum Annealing}
\acrodef{QPU}[QPU]{Quantum Processing Unit}
\acrodef{QUBO}[QUBO]{Quadratic Unconstrained Binary Optimization}

\begin{document}

\title{Analyzing the Effectiveness of Quantum Annealing with Meta-Learning}

\author*[1]{\fnm{Riccardo} \sur{Pellini}}\email{riccardo.pellini@polimi.it}

\author[1]{\fnm{Maurizio} \sur{Ferrari Dacrema}}\email{maurizio.ferrari@polimi.it}

\affil*[1]{
\orgname{Politecnico di Milano},
\orgaddress{
\city{Milan}, 
\country{Italy}}}

\abstract{
The field of Quantum Computing has gathered significant popularity in recent years and a large number of papers have studied its effectiveness in tackling many tasks.
We focus in particular on Quantum Annealing (QA), a meta-heuristic solver for \ac{QUBO} problems.
It is known that the effectiveness of QA is dependent on the task itself, as is the case for classical solvers, but there is not yet a clear understanding of which are the characteristics of a problem that makes it difficult to solve with QA.
In this work, we propose a new methodology to study the effectiveness of QA based on meta-learning models. 
To do so, we first build a dataset composed of more than five thousand instances of ten different optimization problems. We define a set of more than a hundred features to describe their characteristics, and solve them with both QA and three classical solvers. We publish this dataset online for future research.
Then, we train multiple meta-models to predict whether QA would solve that instance effectively and use them to probe which are the features with the strongest impact on the effectiveness of QA. 
Our results indicate that it is possible to accurately predict the effectiveness of QA, validating our methodology. Furthermore, we observe that the distribution of the problem coefficients representing the bias and coupling terms is very informative to identify the probability of finding good solutions, while the density of these coefficients alone is not enough. The methodology we propose allows to open new research directions to further our understanding of the effectiveness of QA, by probing specific dimensions or by developing new QUBO formulations that are better suited for the particular nature of QA. Furthermore, the proposed methodology is flexible and can be extended or used to study other quantum or classical solvers.
}


\maketitle

\section{Introduction}\label{introduction}

In recent years the field of Quantum Computing has gathered significant popularity, thanks to remarkable advancements that led to the development of several quantum computers of different architectures and technologies that can be used to tackle numerous problems. Although quantum computers are still limited both by their relatively small size and by the noise that limits the precision of the computation, the field is rapidly moving forward.
Among the existing Quantum Computing paradigms, Quantum Annealing (QA) is a meta-heuristic that can be used to solve Quadratic Unconstrained Binary Optimization (QUBO) problems, a family of NP-hard optimization problems. 
The key idea of QA is to represent a QUBO problem as an energy minimization problem of a real and configurable quantum device. To do so, the problem variables are mapped onto physical quantum bits, or qubits.
The quantum device is steered towards a state of minimal energy, called \emph{ground state}, with a controlled evolution. The ground state corresponds to the optimal solution of the original QUBO problem. The devices that implement the QA process are called \emph{Quantum Annealers}.

The ability of QA to tackle NP-hard optimization problems and its flexibility to heterogeneous domains is what makes it an interesting technology for industries and researchers. Many applications of QA have been proposed in the fields of machine learning \cite{Neukart2018, Mott2017, Mandra2016, DBLP:conf/sigir/DacremaMN0FC22, Neven2009, WillschWRM20, KumarBTD18, Neukart2018_cluster, OMalleyVAA17, ottaviani2018low, NembriniDC21}, chemistry \cite{Micheletti2021, HernandezA17, streif2019solving, Xia2018} and logistics \cite{Ikeda2019, RieffelVODPS15, Ohzeki2020, StollenwerkLJ17}, but the results are not always competitive against classical heuristics solvers.

An important issue is that the quality of the solutions found by QA is limited by multiple factors. 
First of all, Quantum Annealers are physical devices which have a limited number of qubits and connections between them. This limits the size of the problems that they can tackle and requires to process the QUBO problem adapting it to the physical structure of the Quantum Annealer. 
A second important aspect is that the quality of the solutions found by QA depends on the behaviour of the underlying physical quantum system, which is very difficult to study. It is known that some problems appear to be more difficult to solve with QA \cite{Yarkoni2022, Jiang2023, HuangXLGGW23}, but understanding why is not a trivial task and still an open research question.

Most of the previous studies on QA compare its performance, in terms of required time for computation with respect to other heuristic solvers, rather than on the quality of the solutions it finds, \idest its effectiveness. There are two ways in which one can study the effectiveness of QA, one is by analytically describing the underlying quantum behaviour and the other is to perform empirical experiments.
A theoretical analysis has been performed for very small QUBO instances \cite{Stella2005}, which however are too simple to assess the effectiveness of QA when compared to other classical solvers. Furthermore, analytically analysing such a quantum system becomes rapidly very expensive and is generally impossible for problems of interesting size. 
On the other hand, the existing empirical studies on the effectiveness of QA have explored much larger problems but focus mainly on specific tasks such as feature selection \cite{DBLP:conf/sigir/DacremaMN0FC22}, clustering \cite{KumarBTD18, Neukart2018_cluster} and classification \cite{Mott2017, WillschWRM20, Neven2009}, and therefore lack generality.

To the best of our knowledge, there is no published research which has investigated extensively how the characteristics of the problem impact the effectiveness of QA. 
For this reason, in this study we propose a novel empirical methodology for the analysis of the effectiveness of QA, based on the study of the characteristics of QUBO problems with a meta-learning approach. The general idea consists in generating many QUBO instances, defining a set of features which can describe them and train meta-models to predict whether QA would solve that problem or not. 
Our key contributions are as follows:
\begin{itemize}
    \item The design of an experimental methodology which can be applied to study the effectiveness of QA. This methodology can be used also for other quantum algorithms, such as QAOA \cite{Farhi2014} or VQE \cite{Fedorov2021};
    \item The selection of ten classes of optimization problems, each one with specific characteristics, from which we generate approximately five thousand QUBO instances; 
    \item The design and the generation of a meta-learning dataset, which contains for each of the five thousand instances a selection of a hundred features based on probability theory, statistics and graph theory. We show that using them it is possible to effectively predict whether QA would solve a problem instance effectively or not. We share the meta-learning dataset online for further research;
    \item The analysis of the features of a QUBO problem with the strongest impact on the effectiveness of QA;
\end{itemize}

\section{Background} \label{background}
\subsection{QUBO and Ising Models}\label{background:sub1}
In order to use \acf{QA} to tackle optimization problems, these should be represented with one of two equivalent formulations called \ac{QUBO} and \emph{Ising}, suitable for NP-Complete and some NP-Hard optimization problems \cite{Glover2022,lucas2014}. While the two are equivalent, the QUBO formulation is closer to traditional Operations Research, the Ising formulation is instead closer to Physics.

The objective function in the QUBO model is given by Equation \ref{eq:qubo_problem}, where $x \in \{0,1\}^n$ is a column vector representing the \emph{assignment} of the binary variables $x_1, x_2, ..., x_n$, $n$ is the number of problem variables, $y$ the cost, and $Q \in \mathbb{R}^{n \times n}$ is a real square matrix, either symmetric or upper triangular.
\begin{equation}\label{eq:qubo_problem}
    \min_x y = x^TQx
\end{equation}
We will refer to combinatorial optimization problems written in the QUBO formulation as \emph{QUBO problems}. 
Note that the QUBO formulation does not allow for hard constraints. 
An optimization problem with constraints can be transformed into a QUBO problem by introducing a quadratic penalty term multiplied by a penalty coefficient $p$. The idea is that the hard constraints are transformed in \emph{soft} constraints, such that if they are violated a positive penalty $p$ is added to the cost function making the cost of that variable assignment worse. Note that by using soft constraints we do not have the guarantee that the optimal solution will satisfy the constraints, which may happen frequently if the penalty coefficient $p$ has a value that is too low. 
In general, a quadratic binary optimization problem with equality constraints formulated as $Ax - d = 0$, where $d \in \mathbb{R}^m$ and $A \in \mathbb{R}^{m \times n}$, can be transformed into the following QUBO problem:
\begin{align} \label{eq:penaltyqubo_problem}
    \min_x y &= x^TQx + p \cdot x^TCx \\
    C &= (Ax - d)^T(Ax - d) \nonumber
\end{align}
If the quadratic binary optimization problem also has inequality constraints, those need to be transformed first into equality constraints using \emph{binary slack variables}.
For example, if we have the following constraint:
\begin{equation*}
    x_1 + 2x_2 + 4x_3 \leq 3
\end{equation*}
we can transform it into an equality constraint by introducing the binary slack variables $x_4$ and $x_5$:
\begin{equation*}
    x_1 + 2x_2 + 4x_3 + x_4 + 2x_5 = 3
\end{equation*}
There exist no general rule to choose the best number of slack variables, so multiple strategies can be followed.

A second useful formulation is the \emph{Ising model}, which was developed to describe an energy minimization problem for a system of particles \cite{Glover2022, lucas2014}.
The objective function of the Ising model is given by Equation \ref{eq:ising_problem}, where $s \in \{-1,1\}^n$ is the column vector representing the assignment of the $n$ problem variables $s_1, s_2, ..., s_n$ also called \emph{spin} variables, $J \in \mathbb{R}^{n \times n}$ is the \emph{coupling} matrix that describes the quadratic terms of the objective function and has zero diagonal, $b \in \mathbb{R}^n$ is the bias vector, which contains the linear terms of the objective function. The constant term $c \in \mathbb{R}$ is called offset.
\begin{equation}\label{eq:ising_problem}
    \min_s y = s^TJs + b^Ts + c
\end{equation}
A QUBO problem can be transformed into an Ising problem through a linear mapping of the variables. In particular, a binary variable $x_i$ is transformed into a spin variable $s_i$ according to the following conversion\footnote{This mapping is equivalent to the more commonly used $ x_i = \frac{1+s_i}{2} $ with the difference that in our case a binary $0$ is mapped onto spin $1$ and binary $1$ is mapped onto spin $-1$.}:
\begin{equation*}\label{eq:qubo_spin_mapping}
    x_i = \frac{1-s_i}{2}
\end{equation*}

\subsection{Quantum Annealing and Quantum Annealers}\label{background:QA}
Quantum Annealing (QA) is a meta-heuristic solver for QUBO problems.
It is based on the Adiabatic Quantum Computation (AQC) paradigm, with some relaxations \cite{Yarkoni2022, Morita2008, Farhi2000, Albash2018_AQC, Hauke2020}.
The idea is to represent the optimization problem as an energy minimization one, and then use a configurable device that exhibits the needed quantum behaviour to minimize it. Such a device, the Quantum Annealer, is composed by multiple \emph{qubits} connected between each other. QA works based on a time evolution of the quantum system. The initial state of the system is a default one, easy to prepare, so that the qubits are in a state of minimal energy, \idest the \emph{ground state}. Then, the physical system is evolved slowly over a short amount of time by introducing a dependency on the Ising coefficients of the problem one wishes to solve. This means, for example, slowly changing the magnetic fields the qubits are subject to. At the end of the evolution, the physical system will depend only on the problem and, if the evolution was careful enough, it will still be in the ground state. Since the state of minimal energy is also the solution of the optimization problem, measuring the state of the qubits will yield the values that the problem variables should have. 

The evolution of the system in QA occurs in a noisy environment and is subject to quantum fluctuations, \idest \emph{quantum tunneling}, which helps it explore the solution space. 
The noise of the system and the duration of the evolution influence the results of QA, if the evolution is too fast the system will likely escape its ground state and find a worse solution, while if the evolution is too slow noise may build up and push the system again out of the ground state. 
Due to its stochastic nature, QA acts as a device sampling low-cost solutions in a similar way as other classical solvers do, such as Simulated Annealing.
For this reason, QA is repeated multiple times in order to obtain \emph{samples} of the final state of the quantum system. 

The physical devices that implement QA are called \emph{Quantum Annealers}.
Currently, D-Wave Systems Inc. is the company that provides the Quantum Annealers with the largest number of qubits\footnote{The documentation of the D-Wave Systems' services: \url{https://docs.ocean.dwavesys.com/en/stable/index.html}}. 
For example, the D-Wave Advantage has more than 5000 qubits with a topology called Pegasus, where each qubit is connected to other 15 ones.

Solving a QUBO problem with a Quantum Annealer requires the following steps:
\begin{enumerate}
    \item \textbf{Formulate the problem as a QUBO or an Ising problem}: the coefficients that are needed to configure the Quantum Annealers are those of the Ising formulation, as such the problem needs to be in this form. If the problem has a simpler formulation as a QUBO, the transformation is straightforward. Note that some problems can be formulated as QUBO or Ising easily, while others require more expensive processing.
    
    \item \textbf{Embed the problem on the topology of the device:} since the Quantum Annealer is a physical object, we must fit the problem we want to solve on it, accounting for the limited number of qubits and of the connections between them. 
    This procedure is called \emph{minor-embedding} \cite{Carmesin2022} and maps each problem variable to one or more qubits. If multiple qubits are needed to represent a single problem variable, that is called a qubit \emph{chain}. If the problem has a large number of quadratic terms, a substantial number of qubits may be needed to create all the physical connections.  
    Figure \ref{fig:embedding_example} shows an example of how a simple problem can be mapped on a Quantum Annealer.
    \begin{figure}[ht]
        \centering
        \subfloat[][Problem as a graph $G$ of six variables]{\includegraphics[width=0.45\textwidth]{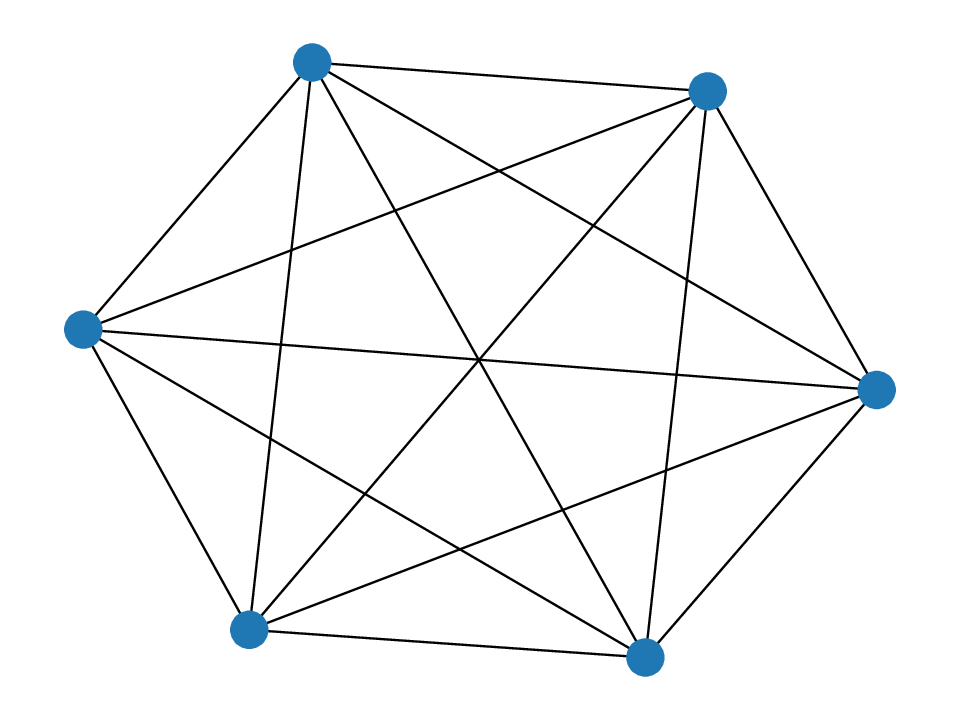}}
        \subfloat[][Embedding of $G$ on Chimera topology]{\includegraphics[width=0.45\textwidth]{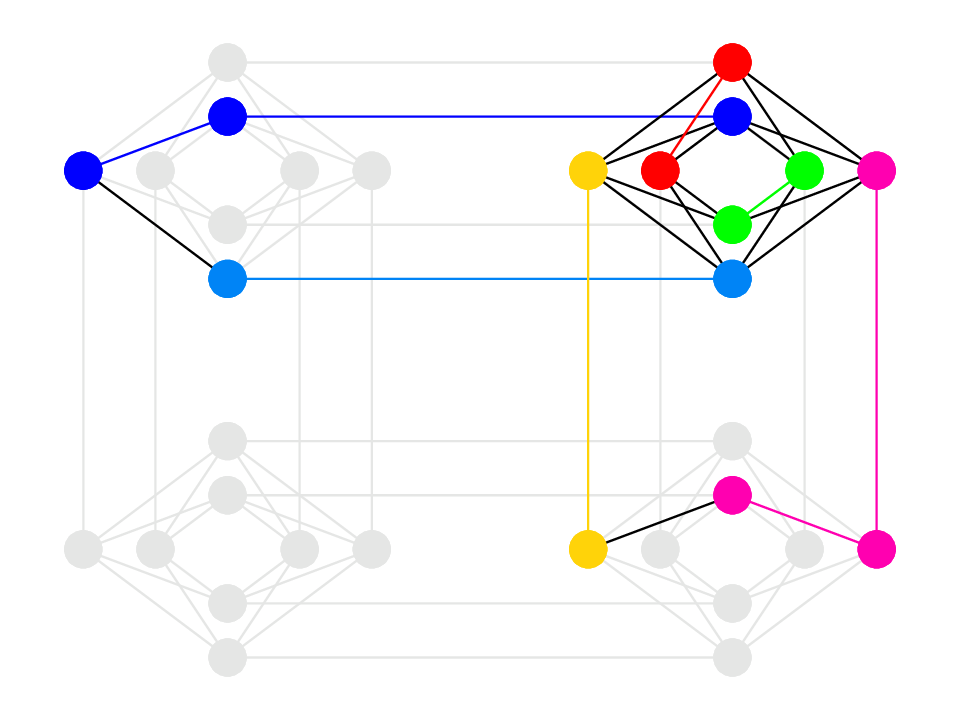}}
        \caption{Embedding of a simple problem with six variables on a portion of a D-Wave Quantum Annealer using the Chimera topology. Each node represents a qubit and each edge a physical connection between them. Nodes of the same colour indicates the chain of qubit used to represent a single problem variable. Note how, while the original problem had six variables, the embedded one requires 14 qubits.
        }
        \label{fig:embedding_example}
    \end{figure}
    Minor-embedding is an NP-Hard problem but polynomial-time heuristic algorithms are available \cite{Choi2008, Cai2014_minorembedding, 
boothby2020nextgeneration}\footnote{We use the library \texttt{minorminer} offered by D-Wave: \url{https://docs.ocean.dwavesys.com/en/stable/docs\_minorminer/source/sdk\_index.html}}.

    \item \textbf{Evolution of the system and sampling of the solutions:}
    once the minor-embedding is done, the problem is transferred to the Quantum Annealer. First, the device is programmed with the problem coefficients, then we can perform a sequence of multiple evolutions to obtain the desired number of samples $n_s$. Each sample requires three steps: (\emph{i}) the evolution is run for the desired duration, called \emph{annealing time} $t_a$, (\emph{ii}) the final state of the system is measured, which requires a read-out time $t_r$ dependent on the number of qubits used, and (\emph{iii}) the device pauses shortly for cooling.
\end{enumerate}

More formally, the energy of a system can be modeled with an \emph{Hamiltonian}, $H \in \mathbb{R}^{2^n \times 2^n}$, and the evolution that occurs in QA is described by the time-dependent Hamiltonian $H(t)$ that models the transition from the initial default Hamiltonian $H_i$\footnote{
{The initial Hamiltonian $H_i$ is also called \emph{driver Hamiltonian}.}} and the Hamiltonian describing the problem $H_p$:
\begin{equation} \label{eq:aqc_hamiltonian}
    H(t) = A(t)H_i + B(t)H_p
\end{equation}
The coefficient $A(t)$ decreases as the evolution progresses, while $B(t)$ increases introducing the dependency on the characteristics of the problem, but their exact values depend on the hardware. At the beginning of the evolution $B(t)$ is zero, while at the end $A(t)$ is zero. Note that this is just a description of the underlying physical system and there is no need to compute this representation to use QA.

In the ideal Adiabatic Quantum Computing setting, it is possible to compute the exact annealing time needed to ensure the system remains in the ground state and finds the global optimum, this result dates back from a century ago \cite{born1928beweis}. This optimal annealing time is inversely proportional to the smallest difference between the two smallest eigenvalues $\lambda_1(t), \lambda_2(t)$ of $H(t)$. Such difference is called \emph{minimum gap}. Although this result may be useful to understand its behaviour, it is not applicable to QA because it is subject to noise. Furthermore, computing the eigenvalues of $H(t)$ is prohibitive for all but the smallest problems.

To exemplify how this representation works, assume to have an Ising problem of $n$ variables with coupling $J$ and bias $b$, $H_p$ is a $2^n \times 2^n$ matrix computed as follows:    \begin{equation} \label{eq:prob_hamiltonian}
    H_p = \sum_{i=1}^n \sum_{j=i}^nJ_{ij}\sigma_z^{(i)}\sigma_z^{(j)} + \sum_i^n h_i \sigma_z^{(i)}
\end{equation}
The matrix $\sigma_z^{(i)}$ is the Z-Pauli operator $\sigma_z$ acting on qubit $i$:
\begin{equation}
    \sigma_z = \begin{pmatrix*}[r]
        1 & 0 \\
        0 & -1
    \end{pmatrix*}
\end{equation}
\begin{equation} \label{eq:sigma_z_i}
    \sigma_z^{(i)} = \bigotimes_{k=1}^{i-1}I  \otimes \sigma_z \bigotimes_{k=1}^{n-i}I \\
\end{equation}
with $\otimes$ being the tensor product and $I$ the identity matrix. A useful property of $H_p$ is that it is a diagonal matrix that contains all the cost values for all possible variable assignments of the problem. Since it is diagonal, these values are also its eigenvalues and the corresponding eigenvectors encode the variable assignment that has that cost. The minimum eigenvalue of $H_p$ corresponds to the minimal cost and the corresponding eigenvector to the optimal variable assignment.

As an example, consider the following QUBO problem, which is minimized when $x_1 = x_2$:
\begin{equation*}
    \min_{x_1, x_2} y = x_{1} + x_{2} - 2x_1 x_2
\end{equation*}
The equivalent Ising formulation is:
\begin{equation*}
    \min_{s_1, s_2} y = \frac{1}{2} - \frac{1}{2}s_1 s_2
\end{equation*}
For this small instance we can compute $H_p$ easily.
The matrices $\sigma_z^{(1)}$ and $\sigma_z^{(2)}$ are:
\begin{equation*}
    \sigma_z^{(1)} = \sigma_z \otimes I = \begin{pmatrix}
        1 & 0 & 0 & 0 \\
        0 & 1 & 0 & 0 \\
        0 & 0 & -1 & 0 \\
        0 & 0 & 0 & -1 \\
    \end{pmatrix}  \quad
    \sigma_z^{(2)} = I \otimes \sigma_z = \begin{pmatrix}
        1 & 0 & 0 & 0 \\
        0 & -1 & 0 & 0 \\
        0 & 0 & 1 & 0 \\
        0 & 0 & 0 & -1 \\
    \end{pmatrix}
\end{equation*}
$H_p$ is then equal to:
\begin{equation*}
    H_{p} = \frac{1}{2}\begin{pmatrix}
        -1 & 0 & 0 & 0 \\
        0 & 1 & 0 & 0 \\
        0 & 0 & 1 & 0 \\
        0 & 0 & 0 & -1 \\
    \end{pmatrix}
\end{equation*}
The smallest eigenvalue of $H_p$ is $\lambda_{\min} = -\frac{1}{2}$ and has a multiplicity of two corresponding to the first and last eigenvalues. Indeed, both $x_0 = 0, x_1=0$ and $x_0 = 1, x_1=1$ are optimal solutions of the problem. If we sum $\lambda_{\min}$ with the offset of the Ising problem, $\frac{1}{2}$, we obtain $0$, that is the same value of the QUBO cost function when $x_1 = x_2$.

\subsection{Studies on the Effectiveness of QA}
Most of the previous studies on QA focus on its performance, by measuring the time required to solve a problem and comparing it to that of classical solvers.
To the best of our knowledge there is no consensus on whether QA provides a general and consistent speedup compared to other traditional solvers for QUBO problems \cite{Hauke2020, Yarkoni2022, Katzgraber2014}, while a recent paper claims substantial speedup for a quantum simulation task \cite{king2024computational}. Although some papers may claim a speedup, this is often based on measurements that only account for part of the process. Indeed, one should consider the time required by all phases: (\emph{i}) formulating the optimization problem as QUBO or Ising, (\emph{ii}) embedding the problem on the QA, (\emph{iii}) sampling the solutions on the device and (\emph{iv}) postprocessing the results if needed (for example by checking if the constraints are satisfied). Frequently, the efficiency of QA is measured by only accounting for the usage of the quantum device itself (programming time and the repeated annealing and read-out) while ignoring the time needed for minor embedding and for creating the QUBO formulation. This gives an incomplete picture of the technology that does not account for two significant bottlenecks. For example, it may be that in a certain situation QA is faster than other traditional methods in solving a specific QUBO problem, but that may not be the case anymore if one includes the minor embedding phase. Furthermore, if it is very computationally expensive to formulate the problem as QUBO, it may be more efficient to use other traditional methods that do not need a QUBO formulation at all.

When comparing the quality of the solutions found by QA and classical solvers, \idest their effectiveness, the published literature usually focuses on problems related to specific fields or even to very specific instances of those problems. Due to this, there is still a limited understanding of how would QA compare in a more general setting. For example, the effectiveness of QA has been analyzed for feature selection \cite{DBLP:conf/sigir/DacremaMN0FC22}, classification \cite{Mott2017, WillschWRM20, Neven2009} and  clustering \cite{Neukart2018_cluster, KumarBTD18}, which are typical machine learning tasks. In the field of chemistry, QA has been applied and analyzed to find the equilibria of polymer mixtures \cite{Micheletti2021}, to find similarities between molecules \cite{HernandezA17} and to find their ground state \cite{streif2019solving}.
The effectiveness of QA in solving problems related to logistics has been analyzed too, for example in solving the Nurse Scheduling Problem \cite{Ikeda2019} and in optimizing the assignments of the gates at the airport \cite{StollenwerkLJ17}.

Previous research also studied the effectiveness of QA from a theoretical perspective by representing analytically the evolution of the time-dependent Hamiltonian $H(t)$ (see Equation \ref{eq:aqc_hamiltonian}) and computing the probability of escaping the ground state \cite{Stella2005}. This approach is however limited by the fact that the size of the Hamiltonian grows exponentially on the number of QUBO problem variables $n$, and the analytical analysis of the Hamiltonian becomes rapidly impractical for all but the smallest problems.
An alternative way is to adopt an empirical approach, by using the outcome of multiple experiments to probe the underlying physical system \cite{Irsigler2021}. The idea is to allow the evolution to progress up to a certain intermediate stage and then drastically accelerating it (\idest a quench, according to D-Wave terminology), observing how the effectiveness changes based on when was the evolution accelerated. While this approach allows to tackle large problem instances, applying the acceleration at different stages of the evolution requires to repeat the experiment a large number of times and therefore this approach too is very resource intensive.

To overcome the limitations of the methods adopted in the literature, this paper proposes a new empirical approach to study how the quality of the solutions found by QA is impacted by the characteristics of the problem.
To achieve this, we first collect a dataset of problem instances belonging to 10 selected problem classes and solve them using both QA and three classical solvers. Then, for each problem instance, we compute a set of features describing various characteristics, from the distribution of the bias coefficients of its  QUBO formulation to the topology of the graph that describes the instance once it has been embedded on the Quantum Annealer. Using this dataset we train a machine learning classifier to identify whether QA was able to find a good solution for that instance and, finally, use it to assess which are the most important problem features.

\section{Meta-Learning Dataset Generation}
In this section, we present the methodology used to generate our meta-learning dataset, on which we train the meta-models to predict the effectiveness of QA. We publish this dataset online for future research.
First, we describe how we select the ten classes of problems we want to solve with QA and the strategies we use to generate the five thousand instances.
Then, we describe how we evaluate the effectiveness of QA, in terms of closeness to the optimal solution of the problem and by comparing QA with other classical methods (Simulated Annealing, Tabu Search and Steepest Descent).
Third, we describe the representations of the QUBO problem we used to compute the approximately one hundred features used to train the meta-models.
Finally, we describe how we solve the instances with QA and with the classical solvers, with a particular focus on the choice of the optimal hyperparameters of the solvers.

\subsection{Selection of Problems and Instances}
We identify a selection of ten different optimization problems that exhibit different characteristics: some have constraints, others do not; some have linear terms, others do not; some have a large number of quadratic terms while others do not, etc. The details on their formulations are reported in Appendix \ref{secA1} and the details on the generation of the instances are in Appendix \ref{appendix:instance_generation}.

The first group contains five classes of optimization problems defined over a graph: Max-Cut, Minimum Vertex Cover, Maximum Independent Set, Max-Clique and Community Detection.
They were selected for the following reasons. Both the Max-Cut and Community Detection problems have a straightforward QUBO formulation that does not require penalties to represent constraints. The Max-Cut, Maximum Independent Set and Minimum Vertex Cover problems share the same quadratic terms in their QUBO matrix, but not the diagonal (\idest the linear terms or bias). The Max-Clique problem is formulated as a Maximum Independent Set problem but defined on the complement graph. The Community Detection problem has a very dense QUBO matrix as there are quadratic terms between all variables and is a relevant problem in Machine Learning \cite{Nembrini2022}.
Since these problems are formulated on a graph, we apply them on four different graph topologies: Erd\"os-Renyi, Cyclic, Star and 2d-grid. Note that in order to have a diversified set of instances we introduce small random perturbations to each topology, consisting in few edge insertions and deletions. The number of insertions and deletions depends on the number of nodes of the graph. More details are reported in Appendix \ref{instance_generation:graph_problems}.

The second group of five optimization problems contains: Number Partitioning, Quadratic Knapsack, Set Packing, Feature Selection and $4 \times 4$-Sudoku. These are a more heterogeneous set than the previous graph-based problems ad so require ad-hoc strategies to generate their instances which we detail in Appendix \ref{instance_generation:no_graph_problems}.
Similarly to the Max-Cut and Community Detection problems, the Number Partitioning problem has a straightforward QUBO formulation with no penalty terms to represent constraints.
Similarly to the Community Detection problem, the Feature Selection problem has a dense QUBO matrix with quadratic terms between all variables.
Finally, the Quadratic Knapsack, Set Packing and $4 \times 4$-Sudoku problems are all Constraint Satisfaction Problems, each with different types of constraints.
In particular, the Quadratic Knapsack problem has inequality constraints that need to be converted in equality constraints using slack variables.

We generate multiple instances of all the problem classes we selected. 
Concerning the size of the problem instances, measured in the number of problem variables, there are two constraints to take into account. First, the D-Wave Quantum Annealer that we use has more than 5000 qubits but, due to their limited connectivity, it is generally possible to tackle problem instances up to between 100-200 variables depending on the structure of the QUBO problem. This is due to the minor-embedding phase. Second, we want the instances to be representative of problems that are not trivial and with a Hamiltonian that could not be analyzed analytically. 
In order to provide a more complete picture, we are also interested to assess the impact of the distribution of the solution space of the problem. This, formally, corresponds to the set of eigenvalues and eigenvectors of the Hamiltonian of the problem $H_p$ (see Section \ref{background:QA}). Unfortunately, it is impractical to compute them for instances of more than 32 problem variables, which may be too small and easy to allow a comparison on the effectiveness of different solvers. For this reasons we decided to create two separate sets of instances:
\begin{itemize}
    \item One set of \emph{large instances}, with 5114 instances of between 69 to 99 variables, the upper range of what can be tackled with the QA; 
    \item One set of \emph{small instances}, with 246 instances of between 27 to 32 variables. With this set of instances we can do a more complete analysis which includes also the distribution of the solution space.
\end{itemize}

The number of problem instances for each optimization problem is summarized in Table \ref{tab:instances}. 
Notice that the instances of the $4 \times 4$-Sudoku problem are included only in the small instances set, since the largest possible instance is unique and it has at most 64 variables.
All the generated instances are satisfiable and, when needed, the penalty term coefficient $p$ used in the QUBO formulation is optimized with a Bayesian Search\footnote{The range of possible values of $p$, for a particular instance, depends on the cost function of that instance. More details are given in Appendix \ref{appendix:instance_generation}} \cite{Victoria2021, snoek2012}, in order to maximise the number of feasible solutions for Simulated Annealing.

\begin{table}[ht]
    \centering
    \begin{tabular}{c|l|cc}
    \toprule
         & \multirow{2}{*}{Problem Class} &  \multicolumn{2}{|c}{Number of Instances} \\
         &  & Small & Large \\
         \midrule
         \multirow{5}{*}{Graph} 
            & Max-Cut                   & 20 & 620 \\
            & Minimum Vertex Cover      & 20 & 619 \\ 
            & Maximum Independent Set   & 20 & 620 \\ 
            & Maximum Clique            & 20 & 620 \\ 
            & Community Detection       & 20 & 620 \\ 
          \midrule
          \multirow{5}{*}{No-Graph} 
            & Quadratic Knapsack        & 20 & 620 \\
            & Set Packing               & 20 & 620 \\ 
            & Number Partitioning       & 20 & 620 \\ 
            & Feature Selection         & 20 & 155 \\ 
            & $4 \times 4$-Sudoku       & 30 & - \\
          \bottomrule
    \end{tabular}
    \caption{Each row of the table gives the number of small and large instances related to each problem class.}
    \label{tab:instances}
\end{table}

\subsection{Evaluating the Effectiveness of a Solver}\label{effectiveness_solver}
In this section, we describe how to evaluate the effectiveness of a solver and, in particular, of QA.
Both QA and the traditional solvers we compare it to are stochastic and are executed multiple times to obtain a set of variable assignments that aim to minimize the cost function, which we call a set of \emph{samples}. 
A sample is represented by an assignment of the decision variables $x$ and by the related \emph{cost value} $y$.

We solve all instances with QA, Simulated Annealing (SA), Tabu Search (TS) and Steepest Descent (SD).
Our definition of how much a solver is \emph{effective} is based on whether it finds samples that meet some quality constraints. While for the small instance set it is possible to compute the global optimum, for the large ones it is not feasible to do so and therefore we define the effectiveness in relative terms with respect to the other solvers.

In particular, we evaluate the effectiveness of QA on the large instances set by comparing its samples with those of the traditional heuristic solvers. We define the samples associated to the best cost value for a solver $S$ as $y^S_{min}$.
An instance $I$ is \emph{QA-over-all} if the best solution found by QA is at least as good as the best one found by SA, SD, TS combined. More formally, if $y^{QA}_{min} \leq \min{\{y^{SA}_{min}, y^{TS}_{min}, y^{SD}_{min}\}}$.
Comparing QA with a pool of multiple solvers results in a stricter evaluation of its effectiveness, but the condition that QA has to be at least as good as all the other solvers combined may be too strict.
For this reason, we also compare QA with each individual solver.
An instance $I$ is \emph{QA-over-S} if the best solution found by QA is at least as good as the one found by solver $S$, hence $y^{QA}_{min} \leq y^S_{min}$.

For the small instances we can perform a deeper analysis of the effectiveness because we can explore the full solution space and find the global optimum. This is in practice done by computing the Hamiltonian of the problem, $H_p$, which is a diagonal matrix enumerating the eigenvalues $\lambda$, sometimes called \emph{energy}, of all the variable assignments. The eigenvalue is equivalent to the cost function $y$ but does not include possible constant offsets $c$ from the Ising formulation, therefore $y = \lambda + c$. 
The variable assignment $x$ associated to an eigenvalue  $\lambda$ can be computed starting from the corresponding eigenvector of $H_p$.
The global optimum of an instance is the assignment $x_{\min}$ corresponding to the minimal eigenvalue of $H_p$, $ \lambda_{\min}$. We will refer to the maximum eigenvalue as $\lambda_{\max}$.

We define a sample with energy $\lambda$ as \emph{$\epsilon$-Optimal} if the following condition holds:
    \begin{equation}\label{eq:epsilon_optimality}
        \lambda \leq \lambda_{\min} + \epsilon \cdot (\lambda_{\max}- \lambda_{\min})
    \end{equation}
The $\epsilon$-Optimality condition describes how close is the eigenvalue of a sample to the solution of the instance.
The coefficient $\epsilon \in [0,1]$ allows to restrict the interval under which $\lambda$ is considered close enough to the optimal eigenvalue $\lambda_{\min}$. Notice that if $\epsilon=0$ only the global optimum of the instance meets the constraint in Equation \ref{eq:epsilon_optimality}. 

We also define a sample $x$ as \emph{Hamming-Optimal} (h-Optimal) if 
{it differs from any solution $x_{min}$ in at most one decision variable}.
This corresponds to check the Hamming distance between a sample and a solution of the instance:
    \begin{equation} \label{eq:w1b_optimality}
        \left|\left|x - x_{\min}\right|\right|_{\text{Hamming}} \leq 1
    \end{equation}

\subsection{Meta-Learning Features}\label{features_QUBO}

In this section, we introduce the features we define to describe a problem instance.
We rely on a selection of metrics used in statistics and probability theory, such as the Gini coefficient \cite{damgaard2000_gini}, the Herfindahl-Hirschman index \cite{Brezina2016_hhi} and the Shannon entropy \cite{shannon_entropy}, as well as metrics used in graph theory, such as the spectral gap, the radius a graph, its diameter and its connectivity. In total we compute 107 features, which we describe in detail in  Appendices \ref{appendix:definition_of_features} and \ref{appendix:domains_and_components}.

The features we compute can be grouped in multiple domains of analysis.
Overall we identify seven domains, among which we describe the three most relevant ones:
\begin{itemize}
    \item \textbf{Logical Ising Graph (LogIsing):} This domain uses the Ising formulation of a QUBO problem. It is represented as a graph having one node per problem variable, associated to the corresponding bias $b$, and an adjacency matrix that corresponds to the coupling matrix $J$;
    \item \textbf{Embedded Ising Graph (EmbIsing):} This domain uses the Ising formulation of a QUBO problem obtained after its minor embedding on the QA. The target architecture is D-Wave Advantage with the Pegasus topology. Therefore, this formulation represents the actual problem solved by the Quantum Annealer, in which multiple qubits may be used to represent one problem variable. This formulation is represented as a graph in the same way as LogIsing; 
    \item \textbf{Solution Space (SolSpace):} This domain uses the eigenvalues, or energy values, of all possible variable assignments, which can be computed only for the small instances, and aims to describe how they are distributed.
\end{itemize}
Other domains we identify are: (\emph{i}) Normalized Multiplicity (NorMul), whose features are related to the multiplicity of the eigenvalues of $H_p$; (\emph{ii}) Matrix Structure (MatStruct), which contains features related to the distribution of the values of the matrix $Q$ of a QUBO problem; (\emph{iii}) 25\%-SolSpace and 25\%-NorMul, which contain the same features of SolSpace and NorMul, but computed by considering only the 25\% lowest eigenvalues of $H_p$, \idest the energies of the 25\% best solutions.

For the LogIsing and EmbIsing domains we compute several features on different mathematical objects, such as the coupling matrix $J$, the Laplacian matrix of the corresponding graph and the bias vector. We call such objects \emph{components}. 

We can also identify sets of features that refer to the same mathematical object, but are computed on different domains. For example, both LogIsing and EmbIsing domains include features computed on the bias. We refer to them as \emph{component sets} and allow us to perform an analysis of the importance of those mathematical objects that is orthogonal to that of the domains. We identified the following component sets: Coupling, Bias,  Laplacian, Structural Adjacency (StructAdj), Structural Laplacian (StructLap), where
StructAdj and StructLap gather features related to the binarized versions of the coupling and Laplacian matrices. 

\subsection{Hyperparameter Optimization of the Solvers}\label{tuning_solvers}
Since the goal of this study is to compare the effectiveness of different solvers, it is essential to ensure that each solver is using the best hyperparameters. Indeed, it is well-known in many fields that comparing methods that are not consistently optimized leads to inconsistent results that cannot be used to draw reliable conclusions \cite{Shehzad23,DBLP:journals/tois/DacremaBCJ21}. The same applies in our case.

We optimize the hyperparameters of each solver (QA, SA, TS, SD) 
{on the instance with the largest minor-embedding on D-Wave Quantum Annealer} for each optimization problem class. The goal is to identify the hyperparameters that will lead the solver to find the variable assignment with the lowest cost $y$. Once the optimal hyperparameters have been found, they are used to solve all instances of the corresponding problem class. We optimize separately the hyperparameters used for the large and small instances sets. 
{To optimize the hyperparameters of the classical solvers we use the standard QUBO formulation while for QA we use the embedded QUBO formulation: we followed this strategy because the embedded QUBO formulation is required only for QA. In this way, we have a fair comparison between different solvers since, for each one of them, we take into account only the necessary steps to solve a QUBO instance.}

\paragraph{Optimal Hyperparameters of Quantum Annealing}
QA has several hyperparameters that can be optimized\footnote{For a detailed explanation of all the hyperparameters of D-Wave Quantum Annealers, refer to the documentation of the devices: \url{https://docs.dwavesys.com/docs/latest/c\_solver\_parameters.html}}, some of which refer to the evolution process as a whole while others allow to fine-tune it at the level of each individual qubit.
The access to the D-Wave Quantum Annealers is limited and for such a large set of instances we have devised a methodology to optimize the hyperparameters we believe are the most important: 
the \emph{annealing time} $t_a$ and the \emph{number of samples} $n_{s}$. In order to perform an efficient optimization within the available resources, we define a fixed computational budget $T$ for each instance. 
Using the default annealing time, $20 \mu s$, and drawing $100$ samples requires, in the worst case, at most  $37\;ms$. In our experiments we allocated $T=70\;ms$ and $T=300\;ms$ per each problem instance. 

The optimization is performed by iterating over $10$ values for $t_a$, approximately equidistant from each other, between $5\;\mu s$ and $200\;\mu s$. Given $t_a$, the number of samples $n_{s}$ is computed as the maximum value allowed within the computational budget $T$, according to Equation \ref{eq:n_reads_constraint}. For $T =70\;ms$, $n_s$ is between 145 and 537 while, for $T =300\;ms$, $n_s$ is between 766 and 2826.
\begin{equation}\label{eq:n_reads_constraint}
    n_{s} = \left \lfloor \frac{T - t_{p}}{t_a + t_{r} + \Delta} \right \rfloor
\end{equation}
The term $t_p \simeq 15\; ms$ is the time needed to program the instances on the Quantum Annealer, $t_{r}$ is the read-out time, needed to read the results of the annealing process, and $\Delta \simeq 20\; \mu s$ is the delay applied after each read-out operation. The read-out time $t_{r}$ is unknown a-priori because it depends on the size of the embedded problem. Based on empirical observation we use $t_{r}=75 \;\mu s$ for small instances and $t_{r} = 150\;\mu s$ for large instances. 
{We choose $t_a$ and the related $n_s$ which provide the sample with the lowest energy. If for multiple pairs ($t_a$, $n_s$) QA finds samples with the lowest energy, we choose the pair with the smallest $t_a$.}

Since the results we obtained when using both computational budgets are very similar, we report those for $T=70\;ms$. The selected hyperparameters are reported in Table \ref{tab:qa_hyperparameters}.
{Notice that the annealing time $t_a$ for the large instances is often smaller than for the small instances. This highlight that, for large instances, a larger $t_a$ does not improve the effectiveness of QA, at least in the range of values we considered and for the number of samples it allows to draw. Such result suggests that QA may require an additional optimization, for example, of the annealing schedule, which is not straightforward and it goes beyond the scope of this study.}

\begin{table}[ht]
    \centering
    \begin{tabular}{l|rr|rr}
    \toprule
         Problem Class &  \multicolumn{2}{c}{Large Instances} &  \multicolumn{2}{c}{Small Instances} \\
         & $t_a \: [\mu s]$ & $n_{s}$ & $t_a \: [\mu s] $ & $n_{s}$ \\
         \midrule
         Max-Cut & 113 & 190 & 200 & 182 \\
         $4 \times 4$-Sudoku & - & - & 200 & 182 \\
         Max-Clique & 157 & 165 & 135 & 234 \\
         Community Detection & 92 & 205 & 113 & 259 \\
         Number Partitioning & 157 & 165 & 200 & 182 \\
         Maximum Independent Set & 113 & 190 & 135 & 234 \\
         Minimum Vertex Cover & 27 & 273 & 200 & 182 \\
         Set Packing & 70 & 224 & 157 & 214 \\
         Feature Selection & 157 & 165 & 27 & 441 \\
         Quadratic Knapsack & 5 & 307 & 135 & 234 \\
         \bottomrule
    \end{tabular}
    \caption{Optimal Hyperparameters for QA for each problem class.}
    \label{tab:qa_hyperparameters}
\end{table}

\paragraph{Optimal Hyperparameters of the Classical Solvers}

The hyperparameters of Simulated Annealing (SA), Tabu Search (TS) and Steepest Descent (SD) are optimized with the following procedure.
For the optimization of these methods we do not use a fixed computational budget because the technology is fundamentally different and, due to the various stages required by QA, it is not trivial to define such a comparison in a way that is \emph{fair}.
First, we fix the number of samples to $n_s = 200$ which is a value comparable to that used for QA. For half of the the large instances QA uses more samples than the classical solvers, while for the remaining half the opposite is true.
We optimize the hyperparameters with a Bayesian Search of 100 iterations \cite{Victoria2021, snoek2012}. The results are available in Appendix \ref{appendix:hyperparameters_solvers}. 
For TS, we optimize the 
number of restarts of the algorithm and the initialization strategy.
For SD, there are no hyperparameters to optimize, except for the number of samples, which we have already set.
For what concerns SA, we optimized the number of \texttt{sweeps}\footnote{A sweep consists in flipping a randomly chosen decision variable.}, the \texttt{schedule}\footnote{The schedule determines how the temperature of the system decreases over time.} and the initial state generator. We noticed however that hyperparameters we found for SA produced worse results compared to the default ones in our following analysis, which may be due to the sensitivity of SA to some of them. For this reason, we retain the default hyperparameters of 1000 \texttt{sweeps}, a \texttt{geometric beta schedule} and a random initial state generator. 

\section{Meta-Model Training and Optimization}\label{validation_metamodels}
In this study we aim to identify which are the characteristics of a problem that impact the effectiveness of QA. We do this by first training a classification model on the dataset we have created in order to predict whether QA would solve that instance well or not based on its features. Since the classifier is trained to predict the outcome on another experiment, it is called a \emph{meta-model}. Once the meta-model is trained, we can use it to probe how important are the various features.

We train the meta-models with Random Forest, AdaBoost, XGBoost and Logistic Regression, using as input data either a specific domain (\eg LogIsing, EmbIsing, SolSpace) or a specific component set (\eg Bias, Coupling, Laplacian), which are described in Section \ref{features_QUBO}.
The target labels are described in Section \ref{effectiveness_solver} (\idest Optimal, $\epsilon$-Optimal, h-Optimal) 
{and they are binary, according to whether the solver meets that effectiveness condition or not.}

The first step is to train the meta-model and optimize its hyperparameters to ensure it is effective in predicting the label. 
In order to measure the effectiveness of the meta-models, we have to account for the significant class imbalance of the labels towards the negative class, \idest instances that are not solved well by QA  (see Section \ref{numerical_results}).
We use Balanced Accuracy ($BA$) to evaluate the meta-models because it is robust to class imbalance.
Given the \emph{true positives} as $TP$, the \emph{true negatives} as $TN$, the number of positive labels in the data as $P$ and the number of negative labels as $N$, the Balanced Accuracy $BA$ is computed as: 
\begin{equation}\label{eq:balanced_accuracy}
    BA = \frac{1}{2}\left(\frac{TP}{P} + \frac{TN}{N}\right)
\end{equation}

The training and optimization of the meta-models is performed with 5-fold Nested Cross-Validation. First, we create a 5-folds \emph{testing split} with a training fold and a testing one which we will use to train and evaluate the meta-model. In order to find the optimal hyperparameters for the meta model, we split each training fold with a further 5-folds split, the \emph{optimization split}. This results in 5 optimization splits for each of the 5 training folds of the testing split and is aimed at preventing the overfitting of the meta-models. 
The splits are all stratified with respect to the problem class of the instances, to ensure every split has an equal distribution of the problem classes. All meta-models are trained on the same data splits and we perform different splits for the large and small instances.
The hyperparameters of the meta-models are optimized according to a Bayesian Search \cite{Victoria2021, snoek2012} exploring 50 configurations, we select those that provide the best Balanced Accuracy on the optimization split.

Once the meta-models have been optimized, we use them to assess which problem characteristic, \idest feature, is most important. We use Permutation Feature Importance (PFI), 
which evaluates how the accuracy of a model drops when the values of a certain feature are shuffled. The idea is that the more important a feature is the larger will be the drop when the values of that features are shuffled. For each feature the process is repeated multiple times and the corresponding importance is given by the mean of the drop in accuracy observed.

\section{Results and Analysis}\label{numerical_results}
In this section we provide the most relevant insights of our analysis regarding the effectiveness of QA.
We have three goals: (\emph{i}) determine hitch classes of
problems are more difficult to solve with QA, (\emph{ii}) understand whether it is possible to predict the effectiveness of QA based on the features we have identified; and (\emph{iii}) discover the domains, the component sets and the features that impact the effectiveness of QA.
To do so, we describe the results obtained by solving the instances with QA and with the other classical solvers. Then, we describe the results of the validation of the meta-models and of the Permutation Feature Importance performed on their features. We publish online a dataset with all the instances we generated, the features we computed and the samples obtained for each solver.\footnote{The instances, the dataset with the features, the results of the solvers and an example script to train meta-models are available at this GitHub repository: \url{https://github.com/qcpolimi/QA-MetaLearning}.}

\subsection{Effectiveness of QA for Large Instances}
\label{sec:effectiveness_QA_large_instances}

\begin{table}[ht]
\centering
\begin{tabular}{lr|rrr}
\toprule
Problem Class & QA-over-all & QA-over-SA & QA-over-TS & QA-over-SD \\
\midrule
Max-Cut                 &       75 \% & 75 \% & 75 \% & 76 \%\\
Number Partitioning     &       30 \% & 30 \% & 33 \% & 30 \%\\
Community Detection     &       13 \% & 15 \% & 24 \% & 39 \%\\
Minimum Vertex Cover    &       0 \% & 77 \% & 0 \% & 0 \%\\
Maximum Independent Set &       0 \% & 73 \% & 0 \% & 0 \%\\
Set Packing             &       0 \% & 32 \% & 0 \% & 0 \%\\
Quadratic Knapsack      &       0 \% & 2 \% & 68 \% & 0 \%\\
Feature Selection       &       0 \% & 0 \% & 0 \% & 0 \%\\
Maximum Clique          &       0 \% & 0 \% & 0 \% & 0 \%\\
\midrule
Average                 &       14 \% & 37 \% & 24 \% & 18 \% \\
\bottomrule
\end{tabular}
\caption{Comparison on the percentage of problem instances in which QA is at least as effective as a specific solver (QA-over-SA, QA-over-TS and QA-over-SD) or as all of them combined (QA-over-all).
}
\label{tab:effectiveness_table_large_instances}
\end{table}

In this section we discuss the effectiveness of QA compared to the other classical solvers (SA, TS and SD) on the large problem instances. Table \ref{tab:effectiveness_table_large_instances} reports the results on each problem class according to the labels we defined in Section \ref{effectiveness_solver}, \idest whether the best sample found by QA is at least as good as that found by a specific solver (QA-over-SA, QA-over-TS and QA-over-SD) or by all of them combined (QA-over-all).

{As a general comment, we can observe that for less than half of the problem classes (4 out of 9) QA solves effectively more instances than at least one classical solver, while for most of the problem classes (7 out of 9) QA is more effective than at least one of the classical solvers for some particular instances.}
However, if we combine all classical solvers QA is more effective only in three problem classes but mostly to a limited extent. Only for Max-Cut QA shows a consistently high effectiveness. 
These results confirm that the effectiveness of QA depends on the problem class, as is the case for classical solvers, which is consistent to what observed in previous studies \cite{Yarkoni2022, Jiang2023, HuangXLGGW23}.
If we compare QA and classical solvers throughout the problem classes, we can see that QA is very frequently more effective than SA, while it is more effective than TS or SD only on some specific problem classes.
As a result, we conclude that comparing QA only with SA, without considering other solvers, is not the best practice to evaluate the effectiveness of QA.

Regarding the characteristics of the problem classes, a first observation we can make is that QA is more effective on problems that do not require penalties to represent constraints: Max-Cut, Community Detection and Number Partitioning. This suggests that the presence of constraints is a factor that makes a problem more difficult to solve with QA. The reason for this may be due to the type of quadratic terms introduced by the penalties which could open new research directions in whether one could use a different formulation for the same constraint that is more suitable for QA \cite{mirkarimi2024experimental}.
Furthermore, remember that Max-Cut, Maximum Independent Set and Minimum Vertex Cover share the same Ising coupling matrix $J$, with the exception of a multiplicative factor, but have a different bias vector $b$. We can observe how Max-Cut is the only one among them that is solved effectively by QA, suggesting that the bias structure plays an important role as well.

Lastly, a high number of quadratic terms (\idest a dense coupling matrix $J$) does not always negatively affect QA. In particular, both Community Detection and Number Partitioning have a dense coupling matrix but still $13\%$ of Community Detection instances and the $30\%$ of Number Partitioning instances are solved effectively with QA.

\subsection{Effectiveness of QA for Small Instances}
\label{sec:effectiveness_small_instances}

In this section we discuss the effectiveness of both QA and the other classical solvers (SA, TS and SD) on the small problem instances. For these instances we can compute the cost associated to all variable assignments and the global optimum. We do this by computing the Hamiltonian of the problem $H_p$ and use its eigenvalues (\idest its diagonal). 
We also compute the maximum energy values needed to assess the $\epsilon$-Optimality for the samples of QA.

\begin{table}[ht]
\centering
\begin{tabular}{l|ccc}
\toprule
Solver &  Optimal &  $10^{-5}$-Optimal &  h-Optimal \\
\midrule
QA &     43 \% &           57 \% &         64 \%  \\
SA &     59 \% &           75 \% &         68 \%  \\
TS &     74 \% &           90 \% &         81 \%  \\
\textbf{SD} &     \textbf{78 \%} &           \textbf{95 \%} &         \textbf{86 \%}  \\
\bottomrule
\end{tabular}
\caption{Fraction of the instances that are solved well according to a certain effectiveness condition (see Section \ref{effectiveness_solver}). The most effective solver is highlighted in bold.}
\label{tab:effectiveness_table_sumup}
\end{table}

Table \ref{tab:effectiveness_table_sumup} compares the effectiveness of the solvers according to the labels defined in Section \ref{effectiveness_solver}, \idest if the solver finds the global optimum (Optimal), if the energy of the best sample is close to that of the global optimum (for $\epsilon$-Optimal we use $\epsilon = 10^{-5}$), if the variable assignment of the best sample has an Hamming Distance of at most 1 with any of the global optimum solutions (h-Optimal).

Consistently with what observed for the large instances, QA is less effective than the classical solvers on all the effectiveness conditions. 
If we compare $10^{-5}$-Optimal and h-Optimal, we can see that QA finds more h-Optimal samples than $10^{-5}$-Optimal samples, as opposed to the other solvers. This indicates that QA finds more easily samples which are close to the optimal ones in terms of Hamming distance rather than energy.
Notice that in this experiment SD is the most effective solver, this may be related to the small size of the instances which may make them relatively easy to solve with simple strategies.

\begin{table}[ht]
\centering
\begin{tabular}{lr|rrr}
\toprule
Problem  Class          &      QA &   SA &   TS &   SD  \\
\midrule
Max-Cut                 &    \textbf{100\%} & \textbf{100\%} & \textbf{100\%} & 88\% \\
Sudoku                  &    \textbf{100\%} & \textbf{100\%} & \textbf{100\%} & \textbf{100\%} \\
Maximum Clique          &    83\% & \textbf{100\%} & 92\% & \textbf{100\%} \\
Community Detection     &    54\% & \textbf{58\%} & \textbf{58\%} & \textbf{58\%} \\
Number Partitioning     &    33\% & \textbf{100\%} & 92\% & \textbf{100\%} \\
Maximum Independent Set &    21\% & 17\% & \textbf{100\%} & \textbf{100\%} \\
Minimum Vertex Cover    &    17\% & 17\% & \textbf{100\%} & 96\% \\
Set Packing             &    12\% & 50\% & 54\% & \textbf{100\%} \\
Feature Selection       &    0\% & \textbf{42\%} & 33\% & 33\% \\
Quadratic Knapsack      &    0\% & 0\% & \textbf{4\%} & 0\% \\
\bottomrule
\end{tabular}
\caption{Fraction of the instances in which the solvers are able to find the global optimum (\idest  Optimal).
The most effective solver of each problem class is highlighted in bold.}
\label{tab:small_effectiveness_table_Optimal}
\end{table}

As done for the large instances, we compare the effectiveness of the solvers on the problem classes. Table \ref{tab:small_effectiveness_table_Optimal} shows the fraction of problem instances in which the solver finds the global optimum (\idest Optimal). Overall, as opposed to what we observed for the large instances, on the small ones QA is never more effective than the classical solvers. 
QA seems to be more effective for problems that are defined over graphs (Max-Cut, Maximum Clique, Community Detection) compared to the ones that are not. Note however that on two graph problems, Minimum Vertex Cover and Maximum Independent Set, QA performs significantly behind the classical solvers. 
Based on the analysis of the large instances (see Section \ref{sec:effectiveness_QA_large_instances}) we observed that the bias of the problem seems to play a role in affecting the effectiveness of QA. This observation is confirmed here as well since QA is much more effective in solving Max-Cut problems than in solving Maximum Independent Set and Minimum Vertex Cover ones.
We can also observe that the effectiveness of QA is quite poor for Set Packing and Feature Selection, being significantly behind the classical solvers. 

Interestingly, on Max-Cut and $4 \times 4$-Sudoku problems almost all the solvers find the global optimum, while for the Quadratic Knapsack hardly any instance can be solved optimally at all, with the most effective solver being TS with 4\% of the instances solved optimally.

As a second analysis we study the effectiveness of QA in sampling solutions with a variable assignment that is close to one of the optimal ones in terms of Hamming distance (\idest h-Optimality). The results reported in Table \ref{tab:effectiveness_table_W1b-Optimal} are consistent with the previous ones in which we assessed the ability of the solvers to find the global optimum, see Table \ref{tab:small_effectiveness_table_Optimal}.
We should note QA exhibits much better effectiveness, when measured in this way, being able to sample solutions close to the optimal ones in the majority of cases. For example, the effectiveness on the Number Partitioning problem goes up from 33\% to 96\% while for Minimum Vertex Cover goes from 17\% to 58\%.
The results also confirm that QA is quite effective for problems that do not require penalties to model constraints (Max-Cut, Community Detection and Number Partitioning). On the other hand, QA is still ineffective for the Feature Selection problem. Quadratic Knapsack remains very challenging for all the solvers.

\begin{table}[ht]
\centering
\begin{tabular}{
lr|rrr}
\toprule
Problem Class &          QA &   SA &   TS &   SD \\
\midrule
Sudoku                  &        \textbf{100\%} & \textbf{100\%} & \textbf{100\%} & \textbf{100\%} \\
Max-Cut                 &        \textbf{100\%} & \textbf{100\%} & \textbf{100\%} & 88\% \\
Number Partitioning     &        96\% & \textbf{100\%} & 92\% & \textbf{100\%} \\
Community Detection     &        96\% & \textbf{100\%} & 96\% & \textbf{100\%} \\
Maximum Clique          &        92\% & \textbf{100\%} & 92\% & \textbf{100\%} \\
Minimum Vertex Cover    &        58\% & 33\% & \textbf{100\%} & 96\% \\
Maximum Independent Set &        46\% & 17\% & \textbf{100\%} & \textbf{100\%} \\
Set Packing             &        38\% & 50\% & 54\% & \textbf{100\%} \\
Feature Selection       &        4\% & 71\% & 67\% & \textbf{75\%} \\
Quadratic Knapsack      &        0\% & 0\% & \textbf{4\%} & 0\% \\
\bottomrule
\end{tabular}
\caption{Fraction of the instances in which the solvers are able to find a variable assignment having a Hamming distance of at most 1 with respect to any optimal solution (\idest  h-Optimal).
The most effective solver of each problem class is highlighted in bold.}\label{tab:effectiveness_table_W1b-Optimal}
\end{table}

\subsection{Meta-Models and Feature Importance Analysis}
\label{subsec:metamodels_and_feature_importance}

\begin{figure}[ht]
    \centering
    \includegraphics[width=0.9\textwidth]{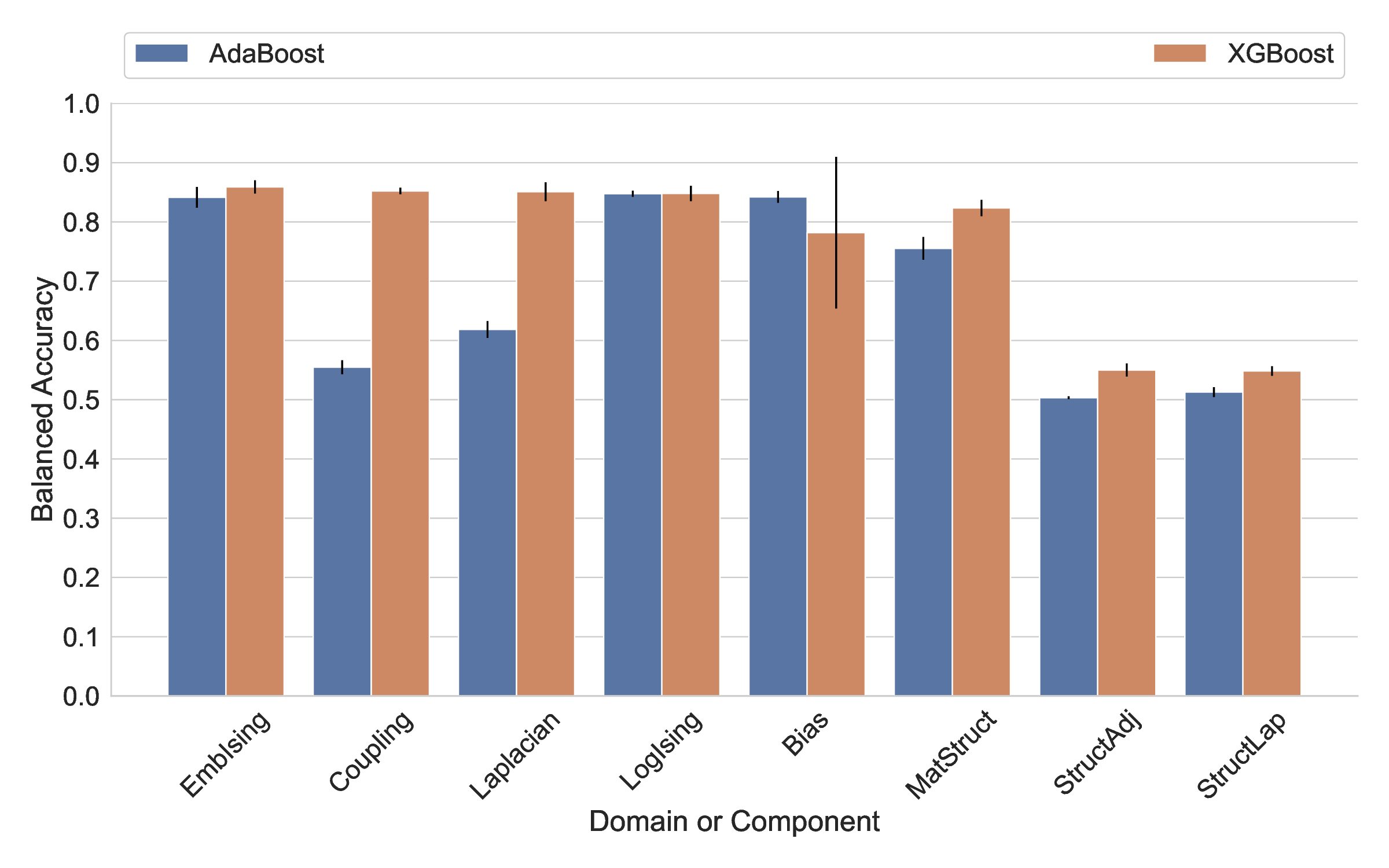}
    \caption{Bar plot showing the Balanced Accuracy of the meta-models which predict whether QA is at least as good of all the classical solvers combined (QA-over-all) on the large instances. The domains or component sets the model is trained on are listed on the x-axis.
    Domains and component sets are ordered according to the Balanced Accuracy of the best related meta-model, in descending order.
    The vertical black segments on the top of each bar represent the standard deviation of the meta-models.}
    \label{fig:score_large_qubo_over_all}
\end{figure}

\begin{figure}
    \centering
    \includegraphics[width=\textwidth]{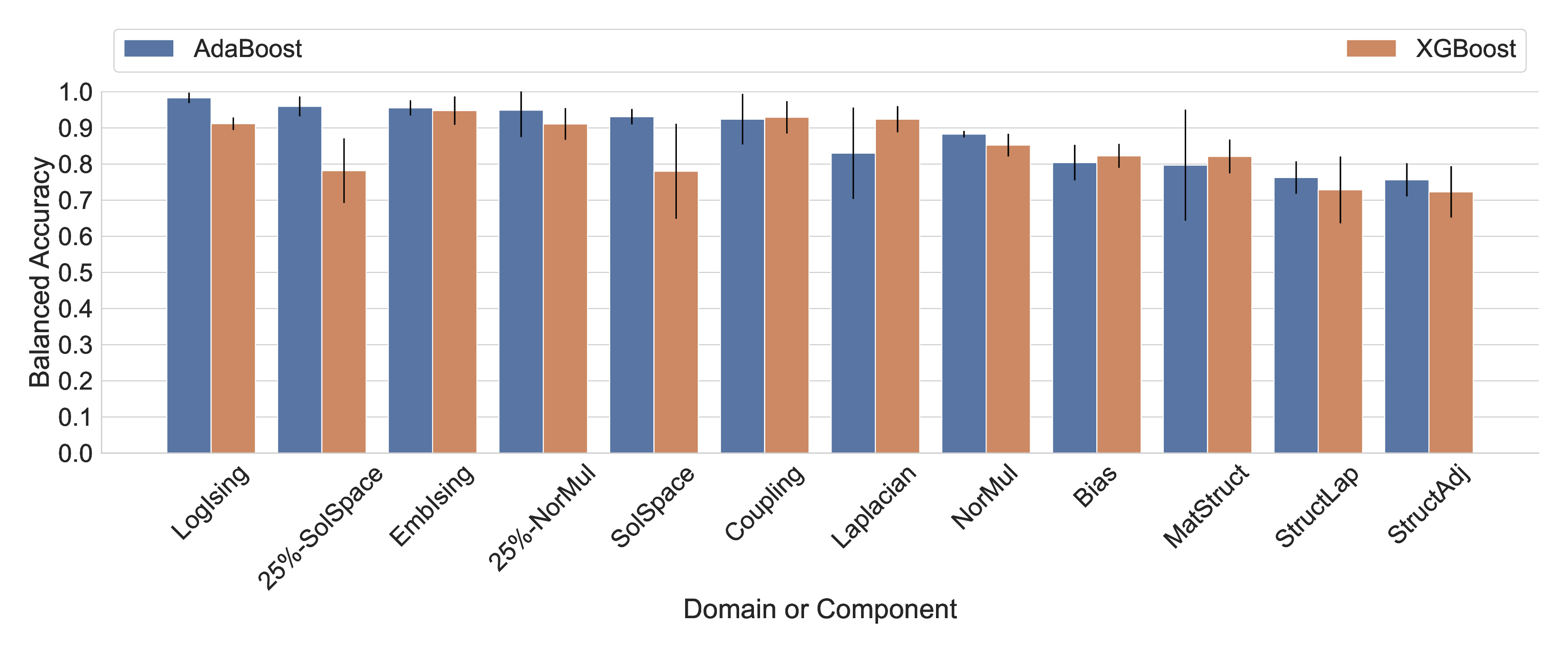}
    \caption{Bar plot showing the Balanced Accuracy of the meta-models which predict whether QA will find the global optimum (Optimal) on small problem instances.
    The domains or components the model is trained on are listed on the x-axis.
    Domains and component sets are ordered according to the Balanced Accuracy of the best related meta-model, in descending ordered.
    The vertical black segments on the top of each bar represent the standard deviation of the meta-models.}
    \label{fig:score_small_qa_optimal}
\end{figure}

The goal of our analysis is to identify the domains and the component sets whose features allow the meta-models to predict well the effectiveness of QA.
We limit our analysis to the meta-models trained to predict whether QA will be at least as effective compared to all the classical solvers combined (target label QA-over-all) for the large instances. We also analyze the meta-models predicting whether QA will find the global optimum on the small instances (QA-Optimal). The full results are available in the online appendix. 

The first important question is whether it is possible to train meta-models able to predict the effectiveness of QA. The results in terms of Balanced Accuracy are shown in Figure \ref{fig:score_large_qubo_over_all} (target QA-over-all) and in Figure \ref{fig:score_small_qa_optimal} (target QA-Optimal) for the two most effective classifiers.
We can immediately see that for several domains or components of the large instances the Balanced Accuracy is approximately 85\%, while for the small instances it is often exceeding 90 or even 95\%. This, combined with the fact that we selected a heterogeneous set of problem classes and instances that are solved by QA with different degrees of success, allows us to conclude that it is indeed possible to build accurate meta-models to predict the effectiveness of QA. These meta-models can then be used for many purposes, among which, studying the behaviour of this technology.

We now analyse which are the domains or components that produce the best meta-models. Concerning the domains, the more informative ones are those related to the graph structure of an Ising problem (LogIsing and EmbIsing) which, based on the high accuracy of the meta-model, we conclude are very informative on the effectiveness of QA. 
Among the domains, the distribution of the values in the Q matrix of the QUBO problem (MatStruct) is less informative, this can be explained by the fact that the problem that is actually solved on the quantum device is represented as Ising and not as QUBO.

Secondly, if we consider the domains related to the distribution of the energies of an Ising problem which are only available for the small instances (SolSpace, NorMul, 25\%-SolSpace, 25\%-NorMul in Figure \ref{fig:score_small_qa_optimal}) it is possible to build meta-models which, again, predict well the effectiveness of QA with a Balanced Accuracy well above 90\%. 
This result shows that the effectiveness of QA also depends on the distribution of the energies of the problem, \idest on how the cost $y$ of the QUBO problem is distributed. Notably, using features based on the solution space allows to achieve comparable Balanced Accuracy with other domains, indicating that both are equally very informative. This is a particularly good result because it is relatively easy to compute the features for the other domains once the problem has been formulated as QUBO.

If we look at the orthogonal grouping of the features, by component sets, we notice that with the Bias, Coupling the Laplacian component sets it is possible to train at least a meta-model with good Balanced Accuracy (higher than $80\%$). On the other hand, the Structural Adjacency (StructAdj) and Structural Laplacian (StructLap) are the least informative and, in particular for large instances, do not allow to build a meta-model better than random guess. Since both StructAdj and StructLap are computed on the binarized problem structure, they only account for how the problem variables are connected and not the coefficient values, this means that the structure of the problem alone is not informative at all.
The bias and the coupling of an Ising problem, together with the Laplacian matrix related to the graph of the Ising problem, are the most informative on the effectiveness of QA.

In general, we can confirm that the characteristics of the problem are important to determine the effectiveness of QA but one must account for the actual coefficient values of the problem and, preferably, use features derived from the Ising formulation. This could open new research questions on whether one can change the formulation of a problem so that its coefficients have a different distribution that is more adequate for the QA. Furthermore, since the coefficients are a function of the problem class and the data, it may be possible to identify which types of graph topologies may be more or less difficult to tackle based on the distributions of the coefficients that they would produce.

We now move to analyzing which specific features are the most important among the ones we identified.
We limit our analysis on feature importance to the XGBoost and AdaBoost meta-models trained on two domains (LogIsing and EmbIsing) and on two component sets (Bias and Coupling).
In the case of meta-models related to small-instances, we include in the analysis also the domain SolSpace.
The best five features of each of these domains and component sets are listed in Table \ref{tab:pfi_table_large_QA_is_best_['ada_boost', 'xg_boost']_QA} (target QA-over-all, large instances) and in Table \ref{tab:pfi_table_small_Optimal_['ada_boost', 'xg_boost']_QA} (target Optimal, small instances).

\begin{table}[ht]
\centering
\begin{tabular}{llll}
\toprule
& {} &  AdaBoost &  XGBoost \\
\midrule
Domains & LogIsing     &                 \texttt{Bias min} &       \texttt{Coupling min eigval} \\
&      &                 \texttt{Bias gini index} &       \texttt{Coupling max eigval} \\
&      &  \texttt{Laplacian connected components} &         \texttt{Degree min eigval} \\
&      &            \texttt{Laplacian min eigval} &           \texttt{Coupling radius} \\
&      &           \texttt{Bias condition number} &           \texttt{Bias gini index} \\
\cmidrule{2-4}
& EmbIsing   &                 \texttt{Bias min} &                  \texttt{Bias hhi} \\
&    &           \texttt{Bias condition number} &  \texttt{Graph Structure qubits} \\
&    &            \texttt{Bias shannon entropy} &     \texttt{Coupling spectral gap} \\
&    &                 \texttt{Bias gini index} &       \texttt{Coupling max eigval} \\
&    &             \texttt{Coupling gini index} &    \texttt{Laplacian spectral gap} \\
\midrule
Component & Bias                    &             \texttt{EmbIsing min} &              \texttt{LogIsing hhi} \\
Sets &                &        \texttt{LogIsing shannon entropy} &       \texttt{EmbIsing gini index} \\
&                     &             \texttt{LogIsing gini index} &       \texttt{LogIsing gini index} \\
&                     &       \texttt{LogIsing condition number} &       \texttt{LogIsing max} \\
&                     &             \texttt{LogIsing min} &  \texttt{LogIsing shannon entropy} \\
\cmidrule{2-4}
& Coupling               &             \texttt{LogIsing gini index} &       \texttt{LogIsing max eigval} \\
&                &                    \texttt{LogIsing hhi} &           \texttt{LogIsing radius} \\
&                &             \texttt{EmbIsing gini index} &           \texttt{EmbIsing radius} \\
&                &             \texttt{LogIsing min eigval} &       \texttt{EmbIsing min eigval} \\
&                &           \texttt{EmbIsing spectral gap} &       \texttt{LogIsing min eigval} \\
\bottomrule
\end{tabular}
\caption{Best five features, ordered according to feature importance, of AdaBoost and XGBoost meta-models trained with LogIsing and EmbIsing domains and with Bias and Coupling component sets.
The target of the meta-models is QA-over-all.}
\label{tab:pfi_table_large_QA_is_best_['ada_boost', 'xg_boost']_QA}
\end{table}

\begin{table}[ht]
\centering
\begin{tabular}{llll}
\toprule
& {} &  AdaBoost &  XGBoost \\
\midrule
Domains & LogIsing     &        \texttt{Coupling gini index} &                \texttt{Coupling max eigval} \\
&     &            \texttt{Bias gini index} &                  \texttt{Degree max eigval} \\
&     &      \texttt{Bias condition number} &                    \texttt{Bias gini index} \\
&    &            \texttt{Coupling radius} &              \texttt{Bias condition number} \\
&     &       \texttt{Structural Adjacency min eigval} &               \texttt{Laplacian max eigval} \\
\cmidrule{2-4}
& EmbIsing   &            \texttt{Bias min} &                \texttt{Coupling max eigval} \\
&   &       \texttt{Bias shannon entropy} &                    \texttt{Bias gini index} \\
&   &            \texttt{Bias gini index} &                           \texttt{Bias hhi} \\
&   &    \texttt{Degree condition number} &  \texttt{Structural Degree shannon entropy} \\
&   &                   \texttt{Bias hhi} &              \texttt{Coupling spectral gap} \\
\cmidrule{2-4}
& SolSpace   &                \texttt{grouped hhi} &               \texttt{gini index} \\
& &       \texttt{gini index} &                        \texttt{grouped hhi} \\
&   &                        \texttt{third quartile} &                                \texttt{third quartile} \\
&   &            \texttt{shannon entropy} &                                \texttt{hhi} \\
&  &                        \texttt{min} &                                \texttt{min} \\

\midrule
Component & Bias                    &  \texttt{EmbIsing condition number} &                \texttt{EmbIsing gini index} \\
Sets &                    &               \texttt{LogIsing hhi} &          \texttt{EmbIsing condition number} \\
&                    &   \texttt{LogIsing shannon entropy} &                       \texttt{LogIsing hhi} \\
&                    &        \texttt{LogIsing min} &           \texttt{EmbIsing shannon entropy} \\
&                    &        \texttt{EmbIsing max} &                       \texttt{EmbIsing hhi} \\
\cmidrule{2-4}
&Coupling               &      \texttt{EmbIsing spectral gap} &              \texttt{LogIsing spectral gap} \\
&               &        \texttt{LogIsing gini index} &              \texttt{EmbIsing spectral gap} \\
&               &      \texttt{LogIsing spectral gap} &                    \texttt{LogIsing radius} \\
&               &          \texttt{LogIsing diameter} &                \texttt{EmbIsing min eigval} \\
&               &        \texttt{EmbIsing min eigval} &                \texttt{EmbIsing gini index} \\
\bottomrule
\end{tabular}
\caption{Best five features, ordered according to feature importance, of AdaBoost and XGBoost meta-models trained on given domains and component sets.
The target of the meta-models is QA-Optimal.
}
\label{tab:pfi_table_small_Optimal_['ada_boost', 'xg_boost']_QA}
\end{table}

\paragraph{Domains Feature Importance}
We consider in particular the domains LogIsing and EmbIsing.
The majority of the most important features are related to the bias and to the coupling of the problem, which is consistent with our previous analysis. Some features are related to the distribution of the values of the bias and the coupling (Gini index, Shannon entropy, Herfindahl-Hirschman index), while other features are related to precise values of these mathematical objects (minimum value, maximum value).

In particular, notice that \texttt{Bias gini index} (related to the distribution of the bias), \texttt{Bias condition number} (related to the values of the bias) and \texttt{Coupling max eigval} (related to the eigenvalues of the coupling) are among the best features in the majority of the meta-models. We deduce that the distribution of the values and the values themselves of the bias are important to study the effectiveness of QA, together with the eigenvalues of the coupling. This is again an interesting observation because it would allow us to identify in advance whether a problem could be well-suited for QA.

Notice also that the number of qubits needed to embed the problem on the Quantum Annealer (\texttt{Graph Structure qubits}) is important, for one meta-model, to predict the effectiveness of QA for the large instances, but not for the small instances. 
This difference may be linked to the fact that, for the small instances, the number of qubits required after the minor-embedding process is limited and therefore has a lower impact.

We analyze, for these two domains, the least important features too. The majority of them is related to the structural adjacency and to the structural Laplacian matrix. This confirms that the sole structure of a problem is not sufficient to determine the effectiveness of QA. Thus, we must consider also the coefficient values between the variables. 

For what concerns the SolSpace domain (see small instances in Table \ref{tab:pfi_table_small_Optimal_['ada_boost', 'xg_boost']_QA}), notice that both the meta-models have the same top three features, although in different order: such features are mostly related to the distribution of the eigenvalues of the problem (\texttt{gini index} and \texttt{grouped hhi}), which plays therefore a role in determining the effectiveness of QA.

\paragraph{Component Sets Feature Importance}
If we consider the Bias component set, the majority of the most important features are related to the distribution of the values of the bias, both considered in the LogIsing domain and in the EmbIsing domain.
In particular, observe that the Gini index and the condition number of the Bias, computed in both domains, are among the five most important features, as they were for the meta-models trained on the domains LogIsing and EmbIsing.
This corroborate our statement that the distribution and the values of the bias are important to determine the effectiveness of QA.

If we instead consider the Coupling component set, we observe that most of the important features we identify are related to the values and to the eigenvalues of the coupling. Features computed in both the LogIsing and EmbIsing domains are important and the most important features are related to the values of the eigenvalues of the coupling.
Notice that the spectral gap and the Gini index of the coupling, computed both in the LogIsing and EmbIsing domains, are shared with almost all meta-models as some of the most important features. We conclude that the eigenvalues of the coupling and their distribution are important to analyze the effectiveness of QA. 

\paragraph{Feature Correlation with Target Label}
In the previous analysis we identified the features that are the most important for the meta-models.
Some of these features, furthermore, are important also if computed in different domains and for different meta-models. These features are, for the LogIsing and EmbIsing domains: \texttt{Bias gini index}, \texttt{Bias condition number}, \texttt{Coupling max eigval}; while for the SolSpace domain:  \texttt{gini index}, \texttt{grouped hhi}, and \texttt{third quartile}. 
We want now to give an intuition of which values of these features determine a low or an high effectiveness of QA.
For each feature we identified, we compute the Spearman rank with the targets of the meta-models (QA-over-all and QA-Optimal).

For most of these features, the Spearman rank does not highlight a strong correlation with the value of the target (in general, the Spearman rank in absolute value is below $0.40$). This suggests that the complexity of the underlying behaviour might require more powerful tools.
Two exceptions are given by \texttt{gini index} and \texttt{grouped hhi} in domain SolSpace, which have respectively a Spearman rank of $-0.596$ and $0.537$.
This mean that as the Gini index of the eigenvalues increases, the Ising problem becomes less difficult to solve.
On the other hand, high values of \texttt{grouped hhi} of the eigenvalues implies that the Ising problem is difficult to solve with QA.

\section{Conclusions}\label{discussion}

In this paper, we have studied the effectiveness of QA with an empirical approach based on meta-learning models.

First, we select a pool of ten optimization problems which can be formulated as QUBO. Then, we generated more than five thousand instances, based on different problem sizes and structures. In particular, we created two sets, one containing large instances and another with small ones for which we can study also the properties of the whole solution space.

As a second step, we define a set of more than a hundred features to describe each problem instance. 
The features are heterogeneous, based on graph theory or on metrics largely used in statistics, probability theory and economics and account for the structure of the problem, its coefficients and its solution space. 
We gather all the features into a meta-learning datasets, which we share on GitHub for further research.

Third, we compare the effectiveness of QA and three classical solvers: Simulated Annealing (SA), Tabu Search (TS) and Steepest Descent (SD).
We observe that QA is frequently less effective than the classical solvers, for both the large and small instances, except for specific problems. 
In particular, we have observed that QA solves more effectively problems with no constraints in their formulation (Max-Cut, Number Partitioning and Community Detection).

Lastly, we train different classification algorithms to predict whether QA will solve an instance effectively or not and show that it is possible to do so accurately. We then use the meta-models to probe the behaviour of QA.
In particular, by analyzing the feature importance of the meta-models, we can observe how the distribution of the bias and the coupling of a problem play a key-role in determining whether QA will be effective in solving it. 

In conclusion, we successfully applied an empirical analysis of the effectiveness of QA based on meta-learning. 
{Possible future directions include the analysis on how different distributions of the coupling and bias values relate to the effectiveness of Quantum Annealing. Such results could be correlated to specific kinds of problems. For example, problems characterized by graphs with a power-law distribution (e.g. problems involving social networks) may be more or less difficult to tackle than those characterized by regular graphs. This can be done, for example, by defining new features which describe how much the distribution of the bias and the coupling differs from another distribution, e. g. from a Gaussian or a uniform distribution.}
Thanks to its generality, the methodology can be easily extended to other heuristic solvers of QUBO problems, such as the Variational Quantum Eigensolver (VQE) \cite{Fedorov2021} and Quantum Approximate Optimization Algorithm (QAOA) \cite{Farhi2014}, providing a useful tool to further our understanding of how to use these quantum algorithms effectively.

\backmatter

\bmhead{Supplementary information}
The meta-learning dataset with all the problem instances, the corresponding graphs and features, as well the samples obtained with each solver can be accessed here: \url{https://github.com/qcpolimi/QA-MetaLearning}.

\section*{Declarations}

This version of the article has been accepted for publication, after peer review but is not the Version of Record and does not reflect post-acceptance improvements, or any corrections. The Version of Record is available online at: \url{https://doi.org/10.1007/s42484-024-00179-8}.

\bmhead{Funding} We acknowledge the financial support from ICSC - ``National Research Centre in High Performance Computing, Big Data and Quantum Computing'', funded by European Union – NextGenerationEU. We acknowledge the CINECA award under the ISCRA initiative, for the availability of high-performance computing resources and support. We also acknowledge the support and computational resources provided by  E4 Computer Engineering S.p.A.

\bmhead{Competing interests} The authors declare no competing interests.

\bmhead{Data availability} The instances, the dataset with the features, the results of the solvers and an example script to train meta-models are available at this GitHub repository: \url{https://github.com/qcpolimi/QA-MetaLearning}.

\bmhead{Author contribution} All authors contributed to the study conception and design.
M. Ferrari Dacrema conceived the methodology.
Material preparation, data collection and analysis were performed by R. Pellini.
The first draft of the manuscript was written by R. Pellini and all authors contributed to the final version. All authors read and approved the final manuscript.

\begin{appendices}

\section{QUBO Formulations of Optimization Problems}\label{secA1}

In this Appendix, we show the QUBO formulations of the ten optimization problems we selected for our study.
We divide the problems in two different groups: problems defined over a graph, explained in Section \ref{subapp:graph_problems}, and problems not defined over a graph, explained in Section \ref{subapp:no_graph_problems}.

\subsection{Graph Problems}
\label{subapp:graph_problems}

\subsubsection{Max-Cut}
Given the graph $G = (V, E)$, a cut over graph $G$ induced by the set of vertices $S$ and $V - S$ is the set of edges which connect vertices in $S$ with vertices in $V-S$.
The Max-Cut problem aims at finding the largest cut which can be induced on graph $G$ \cite{Glover2022}.
Consider, for each  vertex $i$ of graph $G$, a binary variable $x_i$, such that
\begin{equation*}
    x_i = \begin{cases}
        1 \text{ if } i \in S \\
        0 \text{ otherwise}
    \end{cases}
\end{equation*}
Then, the Max-Cut problem is formulated as written in Equation \ref{eq:max_cut_formulation}
\begin{equation}\label{eq:max_cut_formulation}
    \min_x y = -\sum_{(i,j) \in E}(x_i + x_j - 2x_i x_j)
\end{equation}

\subsubsection{Maximum Independent Set}
\label{MIS_formulation}
Given the graph $G = (V, E)$, an independent set is a set $S$ of vertices which are not adjacent to each other.
The Maximum Independent Set problem aims at finding the largest independent set in $G$.
Consider, for each vertex $i$ of graph $G$, a binary variable $x_i$, such that
\begin{equation*}
    x_i = \begin{cases}
        1 \text{ if } i \in S \\
        0 \text{ otherwise}
    \end{cases}
\end{equation*}
Then, the Maximum Independent Set problem can be formulated as a QUBO problem as it is written in Equation \ref{eq:mis_formulation} \cite{Chapuis2018}.
\begin{equation}\label{eq:mis_formulation}
    \min_x y = -a\sum_{i \in V}x_i + b\sum_{(i,j) \in E}x_i x_j
\end{equation}
The penalty term $b$ must outweigh the term $a$, to penalize the choice of having two connected vertices in $S$.

\subsubsection{Minimum Vertex Cover}
Given the graph $G = (V, E)$, a Vertex Cover $C$ is a set of vertices such that all the edges $E$ are connected at least to a vertex in $C$.
The Minimum Vertex Cover is the Vertex Cover with the smallest number of vertices \cite{Glover2022}.
Consider, for each  vertex $i$ of graph $G$, a binary variable $x_i$, such that
\begin{equation*}
    x_i = \begin{cases}
        1 \text{ if } i \in C \\
        0 \text{ otherwise}
    \end{cases}
\end{equation*}
The Minimum Vertex Cover can be formulated as a QUBO problem as written in Equation \ref{eq:mvc_formulation}.
\begin{equation}\label{eq:mvc_formulation}
    \min_x y =  \sum_{i \in V} x_i + p \cdot \sum_{(i,j) \in E}(1 - x_i - x_j + x_i x_j)
\end{equation}
As for the Maximum Independent Set problem, also in the Minimum Vertex Cover problem the penalty term $p$ must be chosen large enough, to penalize the selection of a set of vertices which is not a vertex cover of $G$.

\subsubsection{Max-Clique}
Given the graph $G = (V, E)$, a clique $C$ is a sub-graph of $G$ such that $C$ is fully connected.
The Max-Clique problem aims at finding the clique $C$ of $G$ with the highest number of nodes.
The QUBO formulation of the Max-Clique problem has the same QUBO formulation of the Maximum Independent Set, but it leverages the complement graph $\Bar{G}$ of $G$ \cite{Chapuis2018}.
Please refer to the Section \ref{MIS_formulation}.

\subsubsection{Community Detection}
Given the graph $G = (V, E)$, the aim of the Community Detection problem is to partition $G$ in two communities $C_1, C_2$ of similar nodes \cite{Negre2020, Nembrini2022}.
Remember that the graph $G$ is represented by the adjacency matrix $A$, where the  $(i,j)$ entry is $A_{ij} = 1$ if nodes $i$ and $j$ are connected by an edge, otherwise it is $A_{ij}$.
The degree vector $d$ keeps track of how edges are incident to every node of the graph.
The similarity between nodes is expressed by the \textit{modularity} matrix ${B}$, which is computed according to the following formula:
    \begin{equation*}
        B = A - \frac{{dd}^T}{2|E|}
    \end{equation*}
    The binary variable $x_i$ is related to the node $i \in V$ of graph $G$ and it is defined as follows:
    \begin{equation*}
        x_i = \begin{cases}
            1 & \text{ if } i \in C_1 \\
            0 & \text{ if } i \in C_2
        \end{cases}
    \end{equation*}
    The matrix $Q$ is proportional to ${B}$. In particular, it is equal to:
    \begin{equation*}
    Q = -\frac{1}{|V|} {B}
    \end{equation*}
    The negative sign is due to the fact that QUBO problems require to be \textit{minimization problems} in order to be solved with a QA.
    The formulation of the problem is therefore given by:
    \begin{equation*}
        \min_x y = -\frac{1}{|V|} x^T {B}x
    \end{equation*}

\subsection{No-Graph Problems}
\label{subapp:no_graph_problems}

\subsubsection{Number Partitioning}
Given a set of real numbers $Z = \{z_1, z_2, ..., z_n \}$, 
the Number Partition problem aims at finding two partitions $Z_1, Z_2$ of $Z$ in order to minimize the following expression \cite{Glover2022}:
    \begin{equation}\label{eq:number_partitioning_formula}
        \left(\sum_{z' \in Z_1}z' - \sum_{z'' \in Z_2 }z''\right)^2
    \end{equation}

The binary decision variable $x_i$ is defined as follows:
\begin{equation*}
    x_i = \begin{cases}
            1 & \text{ if } z_i \in Z_1 \\
            0 & \text{ if } z_i \in Z_2
        \end{cases}
\end{equation*}

The QUBO formulation of the Number Partitioning problem is then derived from the expression \ref{eq:number_partitioning_formula}, and it can be written as follows:
\begin{equation*}
    \min_x y = \left( \sum_{i=1}^{|Z|}x_i z_i - \sum_{i=1}^{|Z|}(1-x_i) z_i\right)^2
\end{equation*}
    
\subsubsection{Set Packing}
Given a collection of $n$ sets $S_1, S_2, ..., S_n$, each one with a given capacity $c_1, c_2, ..., c_n$, the Set Packing problem aims at finding a selection of sets providing the largest total capacity, while respecting $m$ constraints for the selection of the sets.
The binary decision variable $x_i$ is associated to the set $S_i$, in particular:
\begin{equation*}
    x_i = \begin{cases}
            1 & \text{ if } S_i \text{ is selected }\\
            0 & \text{ otherwise } 
        \end{cases}
\end{equation*}
The Set Packing problem can be formulated as it follows:
\begin{equation*}
    \begin{aligned}
        \max_x y = & \sum{c_i x_i} \\
        \textbf{s.t.} &  \sum{a_{ji} x_i } \leq 1, \quad j = 1,2,...,m, \quad a_{ji} \in \{0,1\}
    \end{aligned}
\end{equation*}

In order to map the constraints of the Set Packing problem into a quadratic penalty term, we leverage the following conversion rule \cite{Glover2022}:
    \begin{equation*}
    \label{form:set_packing_quadratic_term}
        \sum_{i=1}^n{a_{ji} x_i} \leq 1 \rightarrow p \cdot \sum_{i=1}\sum_{k>i}{a_{ji} a_{jk} x_i x_k}
    \end{equation*}
    where coefficients $a_{ji}$ and $a_{jk}$ are either 0 or 1.
    The objective function is therefore derived as
    \begin{equation*}
        \min_x y = -\sum{c_i x_i} + p \cdot \sum_{i=1}^{n}\sum_{k>i}^{n}{x_i x_k \sum_{j=1}^m a_{ji} a_{jk}}
    \end{equation*}
    
\subsubsection{Quadratic Knapsack}
Given a set of projects $ P_1, P_2, ..., P_n$ such that for each pair $(P_i, P_j)$ it exists a joint revenue $r_{ij} = r_{ji}$, the Quadratic Knapsack problem aims at maximizing the total revenue of activating the projects under a budget constraint. 
The Quadratic Knapsack problem is formulated as it follows:
    \begin{align*}
        \max_x y =  \sum_i^n \sum_j^n{r_{ij} x_i x_j} \\
        \textbf{s.t.} \sum_i^n{c_i x_i }  \leq b
    \end{align*}

    To formulate the Quadratic Knapsack problem in the QUBO formulation, we introduce $m$ binary slack variables $t_1, t_2, ..., t_m$.
    Each slack variable has a budget coefficient $c_{t_j}$, which has to be chosen accordingly to the budget $b$, e. g. by stating that $\sum_j^m c_{t_j} = b$.
    With these type of contraints, we rewrite the problem using the formulation given in Equation \ref{eq:penaltyqubo_problem}, defined in Section \ref{background:sub1}.
    The QUBO formulation of the Quadratic Knapsack problem is therefore:
    \begin{equation*}
        \min_x y = - \sum_i^n \sum_j^n{r_{ij} x_i x_j} + p \cdot \left(\sum_i^n{c_i x_i } + \sum_j^m{c_{t_j} t_j}  - b \right)^2
    \end{equation*}
    
\subsubsection{Sudoku}
A Sudoku can be considered as a constraint satisfaction problem \cite{bukhari2022_sudoku}.
In our study, we analyze only small instances of Sudoku problems, so we consider Sudoku with a $4 \times 4$ grid and some fixed assignments. We call such problems $4 \times 4$-Sudoku
The binary variable we use is defined as
    \begin{equation*}
        x_{(i,j),k} = \begin{cases}
            1  \text{ if cell } (i,j) \text{ has value } k \\
            0 \text{ otherwise }
        \end{cases}
    \end{equation*}
There are four kinds of constraints in a Sudoku:
\begin{itemize}
    \item[] \textbf{\textit{Cell Constraints.}} \\ A cell can contain only one number;
    \begin{equation*}
        \sum_{k = 1}^{4}{x_{(i,j),k}} = 1 \quad \forall i,j \in \{1,2,3,4\}
    \end{equation*}
    \item[] \textbf{\textit{Row Constraints.}} \\ Two cells in the same row must have distinct numbers;
    \begin{equation*}
        \sum_{j = 1}^{4}{x_{(i,j),k}} = 1 \quad \forall i,k \in \{1,2,3,4\}
    \end{equation*}
    \item[] \textbf{\textit{Column Constraints.}} \\ Two cells in the same column must have distinct numbers;
    \begin{equation*}
        \sum_{i = 1}^{4}{x_{(i,j),k}} = 1 \quad \forall j,k \in \{1,2,3,4\}
    \end{equation*}
    \item[] \textbf{\textit{Block Constraints.}} \\ Two cells in the same block must have distinct numbers;
    \begin{align*}
        \sum_{(i,j) \in B}{x_{(i,j),k}} = 1 \quad &
        \forall k \in \{1,2,3,4\}, \\
         & \forall \text{ blocks  B}
    \end{align*}
    There are totally 64 constraints in a $4 \times 4$-Sudoku problem.
    If some cells are already assigned, the number of constraints and the number of variables reduces.
    In particular:
\begin{itemize}
    \item If cell $(i,j)$ is assigned, there exist no variable related to this cell;
    \item If cell $(i,j)$ is not assigned and the number of values it can have is $m$, then there exist $m$ variables related to cell $(i,j)$ and each variable refer to a possible value of the cell;
\end{itemize}

    We can map all the constraints into a single square matrix A, which has at most 64 rows and 64 columns.
    Each column is related to a variable $x_{(i,j), k}$, while each row refers to a constraint.
    Assume that $x$ is a vector where each element is a variable $x_{(i,j), k}$.
    If $\vec{1}$ is a vector of 64 elements, all equal to 1, the QUBO formulation of the $4 \times 4$-Sudoku problem is:
    \begin{equation*}
        \min_x y = (Ax - \vec{1})^T(Ax - \vec{1})
    \end{equation*}
\end{itemize}

\subsubsection{Feature Selection}

Given a dataset $\mathcal{D}$, the Feature Selection problem aims at finding the subset $S \subset F = \{f_1, f_2, ..., f_n\}$  of the best $k$ features able to represent the dataset. This is done especially in machine learning, to reduce the number of features of a predictive model and therefore its complexity.
Assume to have $n$ features. For each feature $f_i$, we have the binary decision variable $x_i$, defined as follows:
\begin{equation*}
    x_i = \begin{cases}
        1 & \text{ if } f_i \text{ is selected} \\
        0 & \text{ otherwise }
    \end{cases}
\end{equation*}
We model the Feature Selection problem using Pearson correlation $Corr(\cdot, \cdot)$ \cite{DBLP:conf/sigir/DacremaMN0FC22}. 
For the quadratic terms, we compute the correlation between two different features $f_i, f_j$ of $\mathcal{D}$.
For the linear terms, we compute the correlation between a feature $f_i$ and the target $t$ of $\mathcal{D}$.
We can compute directly the QUBO matrix $Q$, where in particular its $(i,j)$ element of the matrix $Q$ is equal to:
    \begin{equation*}
        Q_{ij} = 
        \begin{cases}
            Corr(f_i, f_j) & \text{ if } i \neq j \\
            - Corr(f_i, t) & \text{ otherwise }
        \end{cases}
    \end{equation*}
Suppose we want to select $k$ features. To do so, we consider the following QUBO formulation of the Feature Selection problem:
    \begin{equation*}
        \min_x y = x^TQx + \left(\sum_{i=1}^{n}x_i - k\right)^2
    \end{equation*}
Notice that, if less or more than $k$ features are selected, the penalty term is larger than 0 and the objective function value gets worse.

\section{Instance Generation Strategies}
\label{appendix:instance_generation}
In this section, we explain the strategies we applied to generate the instances of the optimization problems we selected.
The strategies varies whether the problem is defined over a graph or not.
An instance of a problem is characterized by its \emph{structure}, which is either the topology of a graph or a general title which refers to the strategy used to generate the constraints and the objective function, and the number of variables $n$.

In general, for all the problems we generate instances with a minimum of $n_{min}$ variables to a maximum of $n_{max}$ variables.
For every optimization problem, except for the $4 \times 4$-Sudoku and for the Feature Selection problem, there exist $n_{rep}$ instances having a certain structure and with the same number of variables.
Such instances are however different between each other, thanks to graph tweaking (explained in Section \ref{instance_generation:graph_problems}) and to the random generation of the coefficient of the cost function and of the constraints.

For the small instances, we have chosen $n_{min} = 27$, $n_{max} = 32$ and $n_{rep} = 1$.
For the large instances, we have chosen $n_{min} = 69$, $n_{max} = 99$ and $n_{rep} = 5$.

\subsection{Graph Problems}
\label{instance_generation:graph_problems}
The instance of a problem defined over a graph is totally determined by the topology of the graph.
The strategy we applied to generate instances of graph problems is essentially divided in two distinct phases: the first is the generation of the graph $G$ of $n$ nodes, according to a certain topology; the second is the generation of an instance, for every graph problem, defined over ${G}$.

We have selected four different graph topologies to generate the graph $G$: the Star topology, the Cycle topology, the 2d-grid topology and the Erd\"os-Rényi topology.
A key step in the generation of $G$ is the random insertion and removal of a limited number of edges, resulting in the \emph{tweaking} of graph $G$. In this way, we generate more graphs starting from a given topology and with $n$ nodes, which however get tweaked in different ways. 
We ensure always that the tweaked graph is connected.
In particular, the graph tweaking process happens according to the following rules:
\begin{itemize}
    \item At most $ n_{ins} = \left \lfloor \frac{n}{6} \right \rfloor$ edges are inserted;
    \item At most $ n_{rem} = \left \lfloor \frac{n}{8} \right \rfloor$ edges are removed;
    \item To perform the tweaking, we randomly modify the adjacency matrix $A$ of the original graph.
    We start the tweaking of $A$ from the first element of the first row, $A_{11}$, and we proceed left-to-right, till the element in the last column and in the last row, $A_{nn}$.
    \item If $A_{ij}=0$ and less than $n_{ins}$ edges have been inserted, insert the edge $(i,j)$ by setting $A_{ij} = 1$ with probability $p_{ins} = 40\%$;
    \item On the contrary, if $A_{ij}=1$ and less than $n_{rem}$ edges have been removed, remove the edge $(i,j)$ by setting $A_{ij} = 0$ with probability $p_{rem} = 30\%$;
    \item If the tweaked graph is not connected, repeat the procedure;
\end{itemize}

For the Minimum Vertex Cover and the Maximum Independent Set problems we have to set the value of the penalty term coefficient $p$.
For the Minimum Vertex Cover, we have set $p = n$, that is to the number of nodes of the graph $G$.
For the Maximum Independent Set, we have set $p = 2n$. 
In both cases, we have that an assignment which violates a constraint highly penalize the objective function and that it is not a solution of the instance.

\subsection{No-Graph Problems}
\label{instance_generation:no_graph_problems}

For what concerns the problems not defined over a graph, to generate an instance we have to generate the matrices related to the objective function and to the constraints. In particular, we want to analyze satisfiable instances.
We discuss now the strategies we used to generate the instances of each problem we selected.

\subsubsection{Number Partitioning}
To generate the instances of the Number Partitioning problem, we have to define the set $Z$ to partition.
For simplicity, we generate only sets of integer numbers.
We selected three probability distribution to generate the set $Z$, in addition to a fourth strategy where $Z$ contains all the numbers between 1 and the number of variables $n$.
The three probability distributions are: (\emph{i}) an uniform distribution between 1 and 99; (\emph{ii}) a geometric distribution with probability of success $p = 0.02$; (\emph{iii}) a Poisson distribution of mean value $\mu = 50$.

\subsubsection{Set Packing}
An instance of the Set Packing problem is determined by the capacities of the $n$ sets $c_1, c_2, ..., c_n$, the coefficients $a_{j,i}$ of the constraints, and the coefficient of the penalty term $p$.

For what concerns the capacities $c_i$, we sample highest possible capacity coefficient, called $c_{max}$, from a uniform distribution between 10 and 39.
This number represents the highest possible capacity coefficient, called $c_{max}$.
Then, we generate the capacities $c_1, c_2, ..., c_n$ of the sets by sampling them from to a uniform distribution between 1 and $c_{max}$.

To generate the coefficients $a_{j,i}$, we consider a matrix $A$ where the element in the $j$-th row and in the $i$-th columns is equal to $a_{j,i}$. Notice that $A$ has always $n$ columns.
We have defined four strategies to generate matrix $A$:
\begin{itemize}
    \item $A$ is rectangular, with $n-1$ rows and the elements $a_{i,i}$ and $a_{i,i+1}$ are equal to 1, for $i = 1,2,...,n-1$. All the other elements are set to zero. We say that this matrix has a \emph{step} structure.
    We provide below, as an example, a matrix of this kind with n = 5:
    \begin{equation*}
        A = \begin{pmatrix}
            1 & 1 & 0 & 0 & 0 \\
            0 & 1 & 1 & 0 & 0 \\
            0 & 0 & 1 & 1 & 0 \\
            0 & 0 & 0 & 1 & 1 \\
        \end{pmatrix}
    \end{equation*}

    \item $A$ is rectangular, with a number of rows $m$ randomly sampled from a uniform distribution between 2 and $\left \lfloor \frac{n}{2} \right \rfloor$. All the columns of $A$ have exactly one element equal to 1. We say that this matrix has \emph{disjoint rows} structure.
    Below, we provide an example of a disjoint rows matrix, with n = 5 and m = 3:
    \begin{equation*}
        A = \begin{pmatrix}
            1 & 0 & 0 & 1 & 0 \\
            0 & 0 & 1 & 0 & 1 \\
            0 & 1 & 0 & 0 & 0 \\
        \end{pmatrix}
    \end{equation*}

    \item $A$ is a square matrix. The elements on the main diagonal are all equal to 1. Furthermore, for each row of $A$, a random off-diagonal element is set to 1 with probability $p=60\%$. We say that this matrix has an \emph{almost diagonal} structure;

    \item $A$ is rectangular, with a number of rows $m$ sampled from a uniform distribution between 10 and $\left \lfloor \frac{n^2}{2}\right \rfloor$.
    The elements of $A$ are randomly generated between 0 and 1, with equal probability. We say that this matrix has a \emph{random} structure;
    
\end{itemize}

Also the penalty term $p$ must be chosen, in order to penalize the assignments which violate the constraints.
We use 100 steps Bayesian optimization to choose $p$, by maximizing the percentage of feasible solutions found by Simulated Annealing in 30 executions.
The range of values of $p$ is determined by the sum $c_{all} = \sum_{i}^n |c_i|$. In particular:
\begin{equation*}
    \left \lfloor \frac{c_{all}}{3} \right \rfloor \leq p \leq c_{all}
\end{equation*}

\subsubsection{Quadratic Knapsack}
In the Quadratic Knapsack problem, to generate the instances we have to choose: (\emph{i}) the values of the joint revenue $r_{ij}$; (\emph{ii}) the values of the coefficients $c_1, c_2, ..., c_n$ of the constraint; (\emph{iii}) the budget $b$; (\emph{iv}) the number $m$ of slack variables $t_i$ to introduce and their constraints coefficients $c_{t_i}$; (\emph{v}) the value of the penalty term coefficient $p$.

We generate the joint revenue values by generating a matrix.
The general joint revenue value $r_{ij}$ is the element in the $i$-th row and in the $j$-th column of a revenue matrix $R$. By definition of the Quadratic Knapsack problem, $R$ is symmetric.
We generate $R$ according to the following four strategies:
\begin{itemize}
    \item $R$ is diagonal. The elements on the diagonal are sampled from a uniform distribution of integer numbers between 15 and 39. We say that the structure of $R$ is \emph{diagonal};

    \item $R$ has the whole diagonal, with some other few random off-diagonal elements.
    The elements on the diagonal of $R$ are sampled from a uniform distribution of integers numbers between 15 and 39. 
    Then, for every column of $R$, we set with probability $40\%$ a random off-diagonal $r_{ij}$ with an integer number sampled from a uniform distribution between 1 and 24; when the element is set, we make the matrix symmetric by dividing $r_{ij}$ by 2 and setting $r_{ji} = r_{ij}$.
    In the case $r_{ji}$ element is set again, when considering column $j$, the previous value of $r_{ji}$ and $r_{ij}$ is overwritten. We say that $R$ has an \emph{almost diagonal} structure.
    An example of almost diagonal $4 \times 4$ matrix is:
    \begin{equation*}
        \begin{pmatrix*}[l]
        17.0 & 13.5 & 0.0 & 16.0 \\
        13.5 & 32.0 & 0.0 & 0.0 \\
        0.0 & 0.0 & 19.0 & 0.0 \\
        16.0 & 0.0& 0.0 & 26.0 \\
    \end{pmatrix*}
    \end{equation*}

    \item The on-diagonal elements of $R$ are non-zero; also the elements immediately on the right and on the left of on-diagonal elements are non-zero; the on-diagonal elements are integer numbers sampled from a uniform distribution between 15 and 39; the off-diagonal elements immediately to the right of on-diagonal elements are sampled from the same distribution; then, the off-diagonal elements are divided by 2 and the matrix is made symmetric, by setting $r_{ij} = r_{ji}$, for every $i, j \in \{1,2,..., n\}$. We say that $R$ has an \emph{enlarged diagonal} structure.
    An example of a $4 \times 4$ enlarged diagonal is:
    \begin{equation*}
        \begin{pmatrix*}[l]
            15.0 & 11.0 & 0.0 & 0.0 \\
            11.0 & 25.0 & 9.5 & 0.0 \\
            0.0 & 9.5 & 31.0 & 17.0 \\
            0.0 & 0.0 & 17.0 & 37.0 \\
        \end{pmatrix*}
    \end{equation*}

    \item $R$ is generated randomly, every element is sampled from the uniform distribution of integer numbers between 15 and 39; then, all the elements below the diagonal are ignored and set to 0; finally, the off-diagonal elements are divided by 2 and the matrix is made symmetric, by setting $r_{ij} = r_{ji}$, for every $i, j \in \{1,2,..., n\}$. We say that $R$ has a \emph{random} structure.
\end{itemize}

The coefficients $c_1, c_2, ..., c_n$ which appear in the constraint are integer numbers sampled from a uniform distribution between 1 and 15.
The budget $b$ is set as $\phi * \sum_{i=1}^n c_i$, where $\phi \in \mathbb{R}$ is sampled from a uniform distribution between 0.20 and 0.70.

We have chosen to introduce four binary slack variables $t_1, t_2, t_3, t_4$ to transform the inequality constraints into equality constraints. The slack variables are respectively associated to the budget coefficients $c_{t_1}, c_{t_2}, c_{t_3}, c_{t_4}$. We choose also the values of these coefficients as it follows:
    \begin{equation}
        c_{t_1} = \left \lfloor \frac{b}{2} \right \rfloor \quad 
        c_{t_2} = \left \lfloor \frac{b}{4} \right \rfloor \quad c_{t_3} = \left \lfloor \frac{b}{8} \right \rfloor \quad c_{t_4} = b - \sum_{i=1}^3 c_{t_i}
    \end{equation}

We use 100 steps Bayesian Search to choose $p$, by maximizing the percentage of feasible solutions found by Simulated Annealing in 30 executions.
The range of the values of $p$ depends on the sum $r_{all} = \sum_{i=1}^n\sum_{j=1}^n|r_{ij}|$.
In particular:
\begin{equation*}
    1 \leq p \leq r_{all}
\end{equation*}

\subsubsection{Sudoku}
To generate $4 \times 4$-Sudoku instances, we generated randomly 30 different solved $4 \times 4$-Sudoku games.
We then iteratively removed the value of a cell, chosen randomly, of the $4 \times 4$-Sudoku game.
Each time we remove a cell, we increase the number of variables of the related $4 \times 4$-Sudoku instance.
We ensure that the generated instance has a number of variables between 27 and 32.

Notice that it is not possible to build instances larger than 64 variables of the $4 \times 4$-Sudoku problem. For this reason, we generate only small instances for this kind of problem.

\subsubsection{Feature Selection}
We generated the instances of the Feature Selection problem starting from the following public datasets\footnote{Datasets are available on the website OpenML: \url{https://www.openml.org}}: \texttt{waveform-5000}, \texttt{SPECTF}, \texttt{spambase}, \texttt{USPS}, \texttt{isolet}, \texttt{gisette}, \texttt{Bioresponse}.
An instance of the Feature Selection problem depends on the dataset $\mathcal{D}$, on the number of features to select $k$ and on the target variable $t$.

Assume we generate $m$ instances with a number of variables $n$, with $n$ bounded between $n_{min}$ and $n_{max}$.
To select the dataset $D$, we use an iterative procedure.
We start from the dataset \texttt{waveform-5000} and we check if it has more than $n = n_{min}$ non-target features: if this condition is true, we select this dataset and we reduce its number of features to $n$, by deleting randomly chosen features; otherwise, we go to the the next dataset, chosen according to the order we used to list them, and we repeat this check. After the last dataset, \texttt{Bioresponse}, we repeat staring from \texttt{waveform-5000}.
After that $m$ features are selected, we continue the procedure by incrementing $n$.
The procedure goes on until we generate $m$ instances with $n = n_{max}$ variables. 

Assume that the dataset $\mathcal{D}$ is the $i$-th dataset selected to generate an instance of $n$ variables, in the iterative procedure, the number of features to select is computed as follows:
\begin{equation*}
    k = \left \lfloor \frac{n}{5} \right \rfloor + i
\end{equation*}

For what concerns the target variable $t$, every dataset we selected has one or multiple target variables. 
Since this formulation can tackle only one target variable $t$, in case of multiple target variables we randomly select one of them and delete all the others.

For small instances, we considered $n_{min} = 27$, $n_{max} = 32$ and $m = 4$, for a total of 24 instances. The number of features $k$ we select is bounded between 5 and 9.
For large instances, we considered $n_{min} = 69$, $n_{max} = 99$ and $m = 5$, for a total of 155 instances. The number of features we select is bounded between 13 and 23.

\section{Definition of the Features}\label{appendix:definition_of_features}

We introduce in this section the definitions of the features we used in our study.
A subset of these features is based on probability theory and statistics, another subset is based on graph theory, and other features are related to the study of the spectrum of matrices.

\paragraph{Gini index}
Given a set $X$ of positive numbers, the Gini index $Gini(Y)$ is a real number which measures the degree of inequality between the values $y \in Y$ \cite{damgaard2000_gini}. It is comprised between 0 (all the values $y \in Y$ are equal) and 1 (only one value of $Y$ is different from 0).
Given an ordered collection of values $Y = \{y_1, y_2, \cdots, y_m\}$, such that $0 \leq y_1 \leq y_2 \leq \cdots \leq y_m$, the Gini index is computed as:
    \begin{equation}
    \label{eq:gini_index}
        g(Y) = \frac{2\sum\limits_{i=1}^{m}iy_i}{m\sum\limits_{i=1}^{m}y_i} - \frac{n+1}{n}
    \end{equation}

\paragraph{Herdindahl-Hirschman index}
Given a set $\Phi$, called \emph{industry}, composed of $m$ firms $F_1, F_2, ..., F_m$, each one characterized by a market share $S(F_i) \in [0,1]$, such that $\sum_{i=1}^m F_i = 1$.
The Herfindahl-Hirschman index (HHI) is a measure used in economics to quantify the competitiveness of an industry with respect to the market share of the firms which compose the industry \cite{Brezina2016_hhi}.
The HHI of inustry $\Phi$ is defined as:
\begin{equation*}
    HHI(\Phi) = \sum_{F \in \Phi}{S(F)}^2
\end{equation*}
Clearly, $0 < HHI(\Phi) \leq 1$. In particular, $HHI(\Phi) = 1$ when there exists only one firm inside industry $\Phi$, that is $\Phi$ is a monopoly.
On the contrary, if we assume that all the firms have equal market share, the market is competitive and, if $m \to +\infty$, we have that $HHI(\Phi) \to 0$. 

If two firms $F_i, F_j$ merge into a larger new firm $F_z$, we have that $S(F_z) = S(F_i) + S(F_j)$. The HHI computed on the new industry, which include $F_z$, increases.

\paragraph{Shannon entropy}
Given a random variable $Y$, distributed according to a certain distribution $p_Y(y)$.
The Shannon entropy of $Y$ measures the level of uncertainty in the distribution $p_Y(y)$.
In the discrete and finite case, assuming that the possible values of $Y$ are $y_1, y_2, ..., y_m$, Shannon entropy is defined as:
\begin{equation*}
    Sh(Y) = -\sum_{i=1}^{m} p_Y(y_i)\log_2 p_Y(y_i)
\end{equation*}
In general, Shannon entropy can be greater than 1.
The maximum value of Shannon entropy occurs when $Y$ is distributed according to a uniform distribution, where all the values of $Y$ are equally probable and none of them can be predicted more easily than the others.

\paragraph{Condition Number}
Given a matrix $M$, the condition number $CN(M)$ measures how close is matrix $M$ to be singular.
If $M$ is symmetric and real, the $CN(M)$ is computed according to the maximal and minimal eigenvalues of $M$, respectively $\lambda_{max}$ and $\lambda_{min}$:
\begin{equation*}
    CN(M) = \left | \frac{\lambda_{max}}{\lambda_{min}} \right |
\end{equation*}

\paragraph{Radius of a Graph}
Given a graph $G(V, E)$, described by the adjacency matrix $A$.
The \emph{radius} of graph $G$ is the largest eigenvalue of $A$ in absolute value.

\paragraph{Diameter of a Graph}
Given a graph $G(V, E)$, the shortest path between two nodes $i$, $j$ is the sequence of edges having minimal cost which connects $i$ and $j$.
The \emph{diameter} of $G$ is the length of the longest shortest path.
If $G$ is not connected, the diameter is not defined.
In the case of negative weights on the edges that form a cycle, the diameter is not defined too.

\paragraph{Spectral Gap of a Graph}
Given a graph $G(V, E)$, described by the adjacency matrix $A$, the \emph{spectral gap} of $G$ the difference, in modulus, between the the two largest eigenvalues of $A$.
Another definition we use is related to the Laplacian $L$ of the graph. In this case, the spectral gap is the smallest non-zero eigenvalue of the Laplacian $L$ related to the graph.

\paragraph{Connectivity of Graph}
Given a graph $G(V, E)$, described by a positive semi-definite Laplacian matrix $L$, the connectivity is equal to the second smallest eigenvalue of $L$.

\paragraph{Connected Components of a Graph}
Given a graph $G(V, E)$, described by a positive semi-definite Laplacian matrix $L$, the number of connected components of the graph is equal to the multiplicity of the eigenvalue 0 of $L$.

\section{Domains and Component Sets}\label{appendix:domains_and_components}

\subsection{Matrices of an Ising Graph}

We call Ising graph a graph having the coupling matrix $J$ of an Ising problem as adjacency matrix. Each node of the graph corresponds to a variable of the Ising problem.
The nodes of an Ising graph has a weight too, determined by the bias vector $b$ of the Ising problem.
In the case the Ising problem is mapped onto the topology of a Quantum Annealer, we call it \emph{embedded Ising graph}; otherwise, we refer to it as \emph{logical Ising graph}.
We call structural adjacency matrix $A$ of an Ising graph the adjacency matrix related to the unweighted version of the Ising graph.

The degree matrix $D_J$ of the Ising graph is computed starting from the coupling $J$, while the structural degree matrix $D_A$ of the Ising graph is computed starting from $A$.

The Laplacian $L$ of an Ising graph is computed according to the following formula:
\begin{equation*}
    L = D - J
\end{equation*}
The structural Laplacian $L_A$ is computed using $A$ instead of $J$ and $D_A$ instead of $D_J$.

The normalized adjacency matrix ${A_N}$ is computed according the following formula:
\begin{equation*}
    {A}_N = D_A^{-\frac{1}{2}} A D_A^{-\frac{1}{2}}
\end{equation*}
The normalized Laplacian ${L}_N$ is instead equal to:
\begin{equation*}
    L_N = D_A^{-\frac{1}{2}} L_A D_A^{-\frac{1}{2}}
\end{equation*}

\subsection{Domains and Related Features}
We list here all the features computed in every domain we consider.
We refer to the definitions of the features given in Appendix \ref{appendix:definition_of_features}.

\subsubsection{Embedding and Logical Ising Graph (EmbIsing and LogIsing)}
In these domain, we gather features related to the embedded and logical Ising graph.
We compute the features starting from the following mathematical objects: bias, coupling, Laplacian, degree, structural adjacency, structural degree, structural Laplacian, normalized adjacency and normalized Laplacian matrices.
In both the domains, we compute the same features, but the considered mathematical objects vary in size and values.
For both the domains, we use the notation \texttt{Object feature} to refer to the feature \texttt{feature}, in lowercase letters, computed on the mathematical object \texttt{Object}.

\paragraph{Coupling}
Given the coupling $J$, we compute:
\begin{itemize}
    \item \texttt{Coupling gini index}: Gini index of the eigenvalues of $J$. Since eigenvalues may be negative, we shift all the eigenvalues by the subtracting to all of them the smallest eigenvalue of $J$;
    \item \texttt{Coupling hhi}: HHI of the eigenvalues of $J$. The share of the eigenvalues is equal to their multiplicity divided by $n$;
    \item \texttt{Coupling shannon entropy}: Shannon entropy of the eigenvalues of $J$. The probability of an eigenvalue is equal to its multiplicity divided by $n$;
    \item \texttt{Coupling condition number}: condition number of $J$;
    \item \texttt{Coupling radius}: radius of the graph described by adjacency matrix $J$;
    \item \texttt{Coupling diameter}: diameter of the graph described by the adjacency matrix $J$. It may be not defined;
    \item \texttt{Coupling spectral gap}: spectral gap of the graph described by the adjacency matrix $J$;
    \item \texttt{Coupling min eigval}: smallest eigenvalue of $J$;
    \item \texttt{Coupling max eigval}: largest eigenvalue of $J$.
\end{itemize}

\paragraph{Bias}
Given the bias $b$, we compute:
\begin{itemize}
    \item \texttt{Bias gini index}: Gini index of the values of $b$. Since some values of $b$ may be negative, the values are all shifted by subtracting the minimal value of $b$;
    \item \texttt{Bias hhi}: HHI of the values of $b$. The share of the values corresponds to their multiplicity divided by $n$;
    \item \texttt{Bias shannon entropy}: Shannon entropy of the values of $b$. The probability of a value corresponds to its multiplicity divided by $n$;
    \item \texttt{Bias min}: the minimal value of $b$;
    \item \texttt{Bias max}: the maximal value of $b$;
    \item \texttt{Bias condition number}: the fraction between the maximal and the minimal value of $b$, in absolute value.
\end{itemize}

\paragraph{Degree}
Given the degree matrix $D$ of an Ising graph, we compute:
\begin{itemize}
    \item \texttt{Degree gini index}: Gini index of the eigenvalues of $D$. Since eigenvalues may be negative, we shift all the eigenvalues by the subtracting to all of them the smallest eigenvalue of $D$;
    \item \texttt{Degree hhi}: HHI of the eigenvalues of $D$. The share of the eigenvalues is equal to their multiplicity divided by $n$;
    \item \texttt{Degree shannon entropy}: Shannon entropy of the eigenvalues of $D$. The probability of an eigenvalue is equal to their multiplicity divided by $n$;
    \item \texttt{Degree min eigval}: minimal eigenvalue of $D$;
    \item \texttt{Degree max eigval}: maximal eigenvalue of $D$;
     \item \texttt{Degree condition number}: condition number of $D$.
\end{itemize}

\paragraph{Laplacian}
Given the Laplacian matrix $L$ of an Ising graph, we compute:
\begin{itemize}
    \item \texttt{Laplacian gini index}: Gini index of the eigenvalues of $L$. Since eigenvalues may be negative, we shift all the eigenvalues by the subtracting to all of them the smallest eigenvalue of $L$;
    \item \texttt{Laplacian hhi}: HHI of the eigenvalues of $L$. The share of the eigenvalues is equal to their multiplicity divided by $n$;
    \item \texttt{Laplacian shannon entropy}: Shannon entropy of the eigenvalues of $L$. The probability of an eigenvalue is equal to its multiplicity divided by $n$;
    \item \texttt{Laplacian min eigval}: minimal eigenvalue of $L$;
    \item \texttt{Laplacian max eigval}: maximal eigenvalue of $L$;
    \item \texttt{Laplacian connectivity}: the second smallest eigenvalue of $L$. We compute it also if $L$ is not positive semi-definite;
    \item \texttt{Laplacian spectral gap}: the smallest non-zero eigenvalue of $L$;
    \item \texttt{Laplacian connected components}: multiplicity of eigenvalue 0 of $L$. We compute it also if $L$ is not positive semi-definite.
\end{itemize}

\paragraph{Structural and Normalized Matrices}
The features computed for the structural adjacency, degree and Laplacian matrices and for the normalized adjacency and Laplacian matrices are the same computed for their weighted counterparts. The main difference is that the structural and the normalized matrices are positive-semidefinite and there is no need to shift the eigenvalues to compute the Gini index.

\paragraph{Graph Structure in EmbIsing Domain}
In the EmbIsing domain, we compute also two features related to the size of the Embedded Ising graph.
In particular, we compute:
\begin{itemize}
    \item \texttt{Graph Structure qubits}: number of nodes of the Embedded Ising graph, which corresponds to the number of qubits used to solve an Ising problem with the Quantum Annealer;
    \item \texttt{Graph Structure chains}: number of chains of nodes in the Embedded Ising graph; a chain represents a single node of the Logical Ising graph, \idest a single variable of the original Ising problem.
\end{itemize}

\subsubsection{Matrix Structure (MatStruct)}
In this domain, we compute the features related to the values of the QUBO matrix $Q$.
In particular, we compute:
\begin{itemize}
    \item \texttt{gini index}: Gini index of the values of $Q$. We subtract the minimal value of $Q$ from all the elements to guarantee that the values are positive;
    \item \texttt{hhi}: HHI of the values of $Q$. The share of an element of $Q$ is equal to its number of occurrences inside $Q$, divided by $n^2$;
    \item \texttt{shannon entropy}: Shannon entropy of the values of $Q$. The probability of an element of $Q$ is equal to its number of occurrences inside $Q$, divided by $n^2$.
\end{itemize}

\subsubsection{Solution Space (SolSpace)}
In this section, we compute features related to the eigenvalues of the Hamiltonian of the problem $H_p$, related to a QUBO problem.
A generic eigenvalue of $H_p$ is $\lambda_i$.
Assume that $\lambda_{min}$ is the minimal eigenvalue of $H_p$. Remember that, if the QUBO problem has $n$ variable, $H_p$ has $2^n$ eigenvalues.
In this domain, we compute:
\begin{itemize}
    \item \texttt{gini index}: Gini index of the eigenvalues of $H_p$. To have all positive values, eigenvalues are shifted by subtracting $\lambda_{min}$ to all of them;
    
    \item  \texttt{hhi}: HHI of the eigenvalues of $H_p$. We use the value 
    \begin{equation}\label{eq:share_eigenvalue}
        S(\lambda_i) = \frac{\lambda_i-\lambda_{min}}{\sum_{j=1}^{2^n}(\lambda_j - \lambda_{min})}
    \end{equation}
    as the share of an eigenvalue $\lambda_i$;

    \item  \texttt{grouped hhi}: HHI of the eigenvalues of $H_p$, where identical eigenvalues are considered together. The share of the eigenvalue $\lambda_i$ is equal to the sum of the shares of the eigenvalues equal to $\lambda_i$, computed according to the formula \ref{eq:share_eigenvalue}.
    
    \item \texttt{shannon entropy}: Shannon entropy of the eigenvalues of $H_p$. We use as probability of eigenvalue $\lambda_i$ the value described in Equation \ref{eq:share_eigenvalue};
    \item \texttt{min}: the minimal eigenvalue of $H_p$;
    \item \texttt{first quartile}: the eigenvalue of the first quartile of $H_p$;
    \item \texttt{median}: the median eigenvalue of $H_p$;
    \item \texttt{third quartile}: the eigenvalue of the third quartile of $H_p$;
    \item \texttt{max}: the maximal eigenvalue of $H_p$.
\end{itemize}

\subsubsection{Normalized Multiplicity (NorMul)}
In this section, we compute features related to the multiplicity of the eigenvalues of the Hamiltonian of the problem $H_p$, related to a QUBO problem. Assume that there are $m$ different eigenvalues of $H_p$.
The multiplicity of the eigenvalue $\lambda_i$ of $H_p$ is equal to $\mu_i$. We call \emph{normalized multiplicity} $\pi_i$ of $\lambda_i$ the value:
\begin{equation*}
    \pi_i = \frac{\mu_i}{\sum_{j=1}^m \mu_j}
\end{equation*}
In this domain, we compute the following features:
\begin{itemize}
    \item \texttt{gini index}: Gini Index of the normalized multiplicities of the eigenvalues of $H_p$;
    \item \texttt{hhi}: HHI of eigenvalues of $H_p$, where we use their normalized multiplicities as shares;
    \item \texttt{shannon entropy}: Shannon entropy of the eigenvalues of $H_p$, where we use their normalized multiplicities as probabilities;
    \item  \texttt{smallest eig}: the normalized multiplicity of the smallest eigenvalue of $H_p$.
\end{itemize}

\subsubsection{25\%-SolSpace and 25\%-NorMul}
The features computed in these domains are the same computed in the domain SolSpace and NorMul.
The only difference is that here, if consider the eigenvalues in ascending order, we consider only the eigenvalues of $H_p$ in the first quartile.

\subsection{Component Sets}

Component sets gather all the features computed on the same mathematical objects, but in different domains.
In these study, all the component sets contain features computed in both the LogIsing and EmbIsing domains.
For every component set, we use the notation \texttt{Domain feature} to refer to the feature \texttt{feature}, written in lowercase letters, computed in the domain \texttt{Domain} on the mathematical object related to the component set.
The component sets we consider are:
\begin{itemize}
    \item \textbf{Bias}: it contains the features related to the bias;
    \item \textbf{Coupling}: it contains the features related to the coupling;
    \item \textbf{Laplacian}: it contains the features related to the Laplacian matrix;
    \item \textbf{StructAdj}: it contains the features related to the structural adjacency matrix;
    \item \textbf{StructLap}: it contains the features related to the structural Laplacian matrix;
\end{itemize}

\subsection{
{Complexity in Features Computation}}
{The major complexity in the computation of the features is given by the computation of the Hamiltonian of the problem, $H_p$.
To do so it is in fact necessary to compute a $2^n$ real-valued vector, with $n$ number of QUBO variables, according to Equation \ref{eq:prob_hamiltonian}. This of course is related only to the small instances.}

{For what concerns the features related to the large instances, the bottleneck lies in computing all the minor-embeddings of the instances: minor-embedding is, indeed, an NP-Hard problem.}

{Once $H_p$ and the minor-embedding of the instances have been computed, all the features are easily computed without any particular complexity
}.

\section{Hyperparameters of the Solvers}\label{appendix:hyperparameters_solvers}
\paragraph{Tabu Search}

We optimize the number of restarts of the algorithm.
The default number of restarts is 100000.
Table \ref{tab:ts_hyperparameters} contains the optimal hyperparameters found for each problem class.

\begin{table}[ht]
    \centering
    \begin{tabular}{l|rr}
    \toprule
         Problem Class &  \multicolumn{2}{c}{Number of restarts} \\
         & Large Instances  & Small Instances \\
         \midrule
         Max-Cut & 523486 &  990687 \\
         $4 \times 4$-Sudoku & - & 72057  \\
         Max-Clique & 977117  & 1174131  \\
         Community Detection & 1415919 &   372449   \\
         Number Partitioning & 1350888  & 871728   \\
         Maximum Independent Set & 869152   & 1436896  \\
         Minimum Vertex Cover & 388887   & 1047039   \\
         Set Packing & 890741 &   1445991   \\
         Feature Selection & 1037380  & 995175   \\
         Quadratic Knapsack & 452422 &  1305536  \\
         \bottomrule
    \end{tabular}
    \caption{Optimal hyperparameters of Tabu Search for each problem class.}
    \label{tab:ts_hyperparameters}
\end{table}

\paragraph{Simulated Annealing}
We optimized the number of \texttt{sweeps} and the \texttt{schedule} type.
The default value of the number of \texttt{sweeps} is 1000, while the defualt \texttt{schedule type} is \texttt{geometric}.
Notice that we solve the instances also with Simulated Annealing with no hyperparameters optimization and we obtained better results.
Table \ref{tab:sa_hyperparameters} contains the optimal hyperparameters found for each problem class.

\begin{table}[ht]
    \centering
    \begin{tabular}{l|rl|rl}
    \toprule
         Problem Class &  \multicolumn{2}{c}{Large Instances} &  \multicolumn{2}{c}{Small Instances} \\
         & \texttt{sweeps} & \texttt{schedule type} & \texttt{sweeps} & \texttt{schedule type} \\
         \midrule
         Max-Cut & 924 & \texttt{linear} & 592 & \texttt{linear} \\
         $4 \times 4$-Sudoku & - & - & 1260 & \texttt{geometric} \\
         Max-Clique & 630 & \texttt{geometric} & 728 & \texttt{geometric} \\
         Community Detection & 1073 & \texttt{linear} & 1033 & \texttt{linear} \\
         Number Partitioning & 1426 & \texttt{geometric} & 819 & \texttt{linear} \\
         Maximum Independent Set & 626 & \texttt{geometric} & 1192 & \texttt{linear} \\
         Minimum Vertex Cover & 1100 & \texttt{geometric} & 1393 & \texttt{linear} \\
         Set Packing & 625 & \texttt{geometric} & 518 & \texttt{geometric} \\
         Feature Selection & 1388 & \texttt{linear} & 576 & \texttt{linear} \\
         Quadratic Knapsack & 503 & \texttt{geometric} & 1133 & \texttt{linear} \\
         \bottomrule
    \end{tabular}
    \caption{Optimal hyperparameters for Simulated Annealing for each problem class.}
    \label{tab:sa_hyperparameters}
\end{table}

\section{Additional Results on the Effectiveness of Solvers}\label{appendix:extended_results_effectiveness}

\begin{table}[ht]
\centering
\begin{tabular}{lr|rrr}
\toprule
 & \multicolumn{4}{c}{$10^{-5}$-Optimal} \\
Solver &            QA &   SA &   TS &   SD \\
Problem                    &               &      &      &      \\
\midrule
Max-Cut                 &          \textbf{100\%} & \textbf{100\%} & \textbf{100\%} & 88\% \\
Sudoku                  &          \textbf{100\%} & \textbf{100\%} & \textbf{100\%} & \textbf{100\%} \\
Maximum Clique          &          83\% & \textbf{100\%} & 92\% & \textbf{100\%} \\
Community Detection     &          \textbf{100\%} & \textbf{100\%} & \textbf{100\%} & \textbf{100\%} \\
Number Partitioning     &          46\% & \textbf{100\%} & 92\% & \textbf{100\%} \\
Maximum Independent Set &          21\% & 17\% & \textbf{100\%} & \textbf{100\%} \\
Minimum Vertex Cover    &          17\% & 17\% & \textbf{100\%} & 96\% \\
Set Packing             &          25\% & 50\% & 58\% & \textbf{100\%} \\
Feature Selection       &          0\% & 71\% & \textbf{75\%} & \textbf{75\%} \\
Quadratic Knapsack      &          62\% & 88\% & 79\% & \textbf{92\%} \\
\bottomrule
\end{tabular}
\caption{Table containing, for each considered problems, the fraction of $10^{-5}$-optimally solved instances for all the solvers. The values in bold text refer to the most effective solvers for a particular problem}
\label{tab:effectiveness_table_1e-05-Optimal}
\end{table}

\end{appendices}

\bibliography{sn-bibliography}

\end{document}